\documentclass[longauth,useAMS,usenatbib]{aa}  

\usepackage{graphicx}
\usepackage{txfonts}
\usepackage{hyperref}
\usepackage{longtable}
\usepackage{appendix}
\begin{document}

   \title{Two temperate Earth-mass planets orbiting\linebreak the nearby star GJ~1002 \thanks{Based [in part] on Guaranteed Time Observations collected at the European Southern Observatory under ESO programme 1102.C-0744. by the ESPRESSO Consortium, and on observations collected at Centro Astron\'omico Hispano en Andaluc\'\i a (CAHA) at Calar Alto, operated jointly by Junta de Andaluc\'\i a and Consejo Superior de Investigaciones Cient\'\i ficas (IAA-CSIC).}' \thanks{This work makes use of observations from the LCOGT network.}' \thanks{The data used in this paper is available in electronic form at the CDS via anonymous ftp to cdsarc.u-strasbg.fr (130.79.128.5) or via http://cdsweb.u-strasbg.fr/cgi-bin/qcat?J/A+A/}}

   \author{A.~Su\'{a}rez~Mascare\~{n}o\inst{1,2} \and
          E.~Gonz\'alez--\'Alvarez \inst{3} \and
          M.~R.~Zapatero~Osorio \inst{3} \and 
          J.~Lillo-Box \inst{4} \and
          J.~P.~Faria \inst{5,6} \and          
          V.~M.~Passegger \inst{1,2,7,8} \and 
          J.~I.~Gonz\'alez~Hern\'andez \inst{1,2} \and 
          P.~Figueira \inst{9,5} \and 
          A.~Sozzetti \inst{10} \and 
          R.~Rebolo\inst{1,2,11} \and 
          F.~Pepe \inst{9} \and 
          N.~C.~Santos\inst{5,6} \and 
          S.~Cristiani \inst{12} \and 
          C.~Lovis \inst{9} \and           
          A.~M.~Silva \inst{5, 6} \and
          I.~Ribas \inst{13,14} \and 
          P.~J.~Amado \inst{15} \and 
          J.~A.~Caballero \inst{4} \and 
          A.~Quirrenbach \inst{16} \and 
          A.~Reiners  \inst{17} \and       
          M.~Zechmeister \inst{17} \and
          V.~Adibekyan \inst{5,6}\and
          Y.~Alibert \inst{18} \and
          V.~J.~S.~B\'ejar \inst{1,2} \and
          S.~Benatti \inst{19} \and
          V.~D'Odorico \inst{12, 20} \and
          M.~Damasso \inst{10}  \and 
          J.~-B.~Delisle \inst{9} \and   
          P.~Di~Marcantonio \inst{12}
          S.~Dreizler \inst{17} \and
          D.~Ehrenreich \inst{9,21} \and
          A.~P.~Hatzes \inst{22} \and
          N.~C.~Hara \inst{9} \and
          Th.~Henning \inst{23} \and
          A.~Kaminski \inst{16} \and
          M.~J.~L\'opez--Gonz\'alez \inst{15} \and
          C.~J.~A.~P.~Martins \inst{5,24} \and
          G.~Micela \inst{19} \and
          D.~Montes \inst{25} \and 
          E.~Pall\'e \inst{1,2} \and 
          S.~Pedraz \inst{26} \and 
          E.~Rodr\'iguez \inst{15} \and 
          C.~Rodr\'iguez--L\'opez \inst{15} \and
          L.~Tal--Or \inst{27} \and
          S.~Sousa \inst{5,6} \and 
          S.~Udry \inst{9}
          }

   \institute{
   Instituto de Astrof\'{i}sica de Canarias, E-38205 La Laguna, Tenerife, Spain\\
              \email{asm@iac.es}            \and           
Departamento de Astrof\'{i}sica, Universidad de La Laguna, E-38206 La Laguna, Tenerife, Spain \and 
Centro de Astrobiolog\'{i}a (CSIC-INTA), Carretera de Ajalvir km 4, E-28850 Torrej\'on de Ardoz, Madrid, Spain \and
Centro de Astrobiolog\'{i}a (CSIC-INTA), ESAC, Camino Bajo del Castillo s/n, E-28692 Villanueva de la Cañada, Madrid, Spain \and
Instituto de Astrof\'{i}sica e Ci\^encias do Espaço, Universidade do Porto, CauP, Rua das Estrelas, 4150-762 Porto, Portugal \and
Departamento de F\'{i}sica e Astronomia, Faculdade de Ci\^encias, Universidade do Porto, Rua Campo Alegre, 4169-007 Porto, Portugal \and 
Hamburger Sternwarte, Gojenbergsweg 112, D-21029 Hamburg, Germany \and 
Homer L. Dodge Department of Physics and Astronomy, University of Oklahoma, 440 West Brooks Street, Norman, OK 73019, United States of America \and
D\'epartement d’astronomie de l’Universit\'e de Gen\`eve, Chemin Pegasi 51, 1290 Versoix, Switzerland \and
INAF – Osservatorio Astrofisico di Torino, Via Osservatorio 20, 10025 Pino Torinese, Italy \and 
Consejo Superior de Investigaciones Cient\'{i}ficas, 28006 Madrid, Spain \and
INAF – Osservatorio Astronomico di Trieste, Via Tiepolo 11, 34143 Trieste, Italy \and
Institut de Ci\`encies de l’Espai (ICE, CSIC), Campus UAB, c/ Can Magrans s/n, E-08193 Bellaterra, Barcelona, Spain \and 
Institut d’Estudis Espacials de Catalunya (IEEC), c/ Gran Capit\`a 2-4, E-08034 Barcelona, Spain \and
Instituto de Astrof\'{i}sica de Andaluc\'{i}a (IAA-CSIC), Glorieta de la Astronom\'{i}a s/n, E-18008 Granada, Spain \and
Landessternwarte, Zentrum f\"ur Astronomie der Universit\"at Heidelberg, K\"onigstuhl 12, D-69117 Heidelberg, Germany \and 
Institut für Astrophysik und Geophysik, Georg-August-Universit\"at, Friedrich-Hund-Platz 1, D-37077 G\"ottingen, Germany \and
Physikalisches Institut, Universit\"at Bern, Siedlerstrasse 5, 3012	Bern, Switzerland \and 
INAF - Osservatorio Astronomico di Palermo, Piazza del Parlamento 1, 90134 Palermo, Italy \and
Scuola Normale Superiore, Piazza dei Cavalieri 7, I-56126, Pisa,
Italy \and
Centre Vie dans l'Univers, Facult\'{e} des sciences de l'Universit\'e de Gen\`eve, Quai Ernest-Ansermet 30, 1205 Geneva, Switzerland \and
Th\"uringer Landessternwarte Tautenburg, Sternwarte 5, 07778 Tautenburg, Germany \and
Max Planck Institute for Astronomy, K\"onigstuhl 17, D-69117 Heidelberg, Germany \and 
Centro de Astrof\'{\i}sica da Universidade do Porto, Rua das Estrelas,
4150-762 Porto, Portugal \and
Departamento de F\'isica de la Tierra y Astrof\'isica and IPARCOSUCM (Instituto de F\'isica de Part\'iculas y del Cosmos de la UCM) \and 
Centro Astron\'omico Hispano en Andaluc\'ia, Observatorio de Calar Alto, Sierra de los Filabres, 04550 G\'ergal, Almer\'ia, Spain \and 
Department of Physics, Ariel University, Ariel 40700, Israel
}

   \date{Written March-September 2022}


  \abstract
 {We report the discovery and characterisation of two Earth-mass planets orbiting in the habitable zone of the nearby M-dwarf GJ~1002 based on the analysis of the radial-velocity (RV) time series from the ESPRESSO and CARMENES spectrographs. The host star is the quiet M5.5~V star GJ~1002  (relatively faint in the optical, $V \sim 13.8$ mag, but brighter in the infrared, $J \sim 8.3$ mag), located at 4.84 pc from the Sun.
 
 We analyse 139 spectroscopic observations taken between 2017 and 2021. We performed a joint analysis of the time series of the RV and full-width half maximum (FWHM) of the cross-correlation function (CCF) to model the planetary and stellar signals present in the data, applying Gaussian process regression to deal with the stellar activity. 
 
 We detect the signal of two planets orbiting GJ~1002. GJ~1002~b is a planet with a minimum mass $m_p \sin i $  of 1.08 $\pm$ 0.13 M$_{\oplus}$ with an orbital period of 10.3465 $\pm$ 0.0027 days at a distance of 0.0457 $\pm$ 0.0013 au from its parent star, receiving an estimated stellar flux of 0.67 $F_{\oplus}$. GJ~1002 c is a planet with a minimum mass $m_p \sin i $  of 1.36 $\pm$ 0.17 M$_{\oplus}$ with an orbital period of 21.202 $\pm$ 0.013 days at a distance of 0.0738 $\pm$ 0.0021 au from its parent star, receiving an estimated stellar flux of 0.257 $F_{\oplus}$. We also detect the rotation signature of the star, with a period of 126 $\pm$ 15 days. We find that there is a correlation between the temperature of certain optical elements in the spectrographs and changes in the instrumental profile that can affect the scientific data, showing a seasonal behaviour that creates spurious signals at periods longer than $\sim$ 200 days.
 
 GJ~1002 is one of the few known nearby systems with planets that could potentially host habitable environments. The closeness of the host star to the Sun makes the angular sizes of the orbits of both planets ($\sim$ 9.7 mas and $\sim$ 15.7 mas, respectively) large enough for their atmosphere to be studied via high-contrast high-resolution spectroscopy with instruments such as the future spectrograph ANDES for the ELT or the LIFE mission. 
} 
 

   \maketitle
%
\section{Introduction}

The search for potentially habitable Earth-like planets is one of the most exciting endeavours in the field of exoplanets. Thanks to radial-velocity (RV) and transit surveys, over 5000 exoplanets have been discovered\footnote{https://exoplanetarchive.ipac.caltech.edu; October 2022}. The first exoplanets detected were, for the most part, giant planets \citep{Mayor1995, Noyes1997}, as was predicted to happen  \citep{Struve1952}. Then, our technical capabilities rapidly expanded, allowing us to reach increasingly lower mass and smaller radius planets \citep{Santos2004,Bonfils2005, Udry2006} and eventually enter the realm of Earth-mass planets \citep{Pepe2011, AngladaEscude2016, Masca2017c, Faria2022}, but only at short orbital periods. Ninety-five percent of the planets with measured masses < 2 M$_{\oplus}$ orbit with periods shorter than 25 days$^{1}$. All but one of the planets detected via RV, with masses < 2 M$_{\oplus}$, orbit with periods shorter than 20 days$^{1}$.

Currently we know a few tens of exoplanets with masses similar to that of the Earth (67 planets with mass or $m_p \sin i $ < 2 M$\oplus$$^{1}$), and hundreds with radii comparable to that of the Earth (855 planets with radius < 1.5 R$\oplus$$^{1}$). However, the number of known exoplanets in the habitable zones (HZs) of their parent stars (i.e. the region in which liquid water can exist in the surface of the planet) with prospects for atmospheric characterisation remains very small. The TRAPPIST-1 system \citep{Gillon2016} is currently the best candidate for atmospheric characterisation via transmission spectroscopy. Another possible path for atmospheric characterisation is high dispersion coronagraphy (HDC) spectroscopy \citep{SparksFord2002, Lovis2017, Blind2022}. This technique allows to combine high-contrast imaging and high-resolution spectroscopy to study the reflected light of the planet. Nearby planets, such as Proxima~b \citep{AngladaEscude2016} or GJ~514~b \citep{Damasso2022}, are ideal candidates for this approach.

Nearby low-mass M-dwarfs offer a unique opportunity in the search of characterisable exoplanets in their HZs. Their low masses allow for the detection of planets of masses similar to that of the Earth without the need of a RV precision as extreme as that required to detect these planets orbiting solar-type stars. Their low luminosity moves the HZ closer to the star, making it possible to sample several orbits within the HZ in modest baselines. The slow rotation of the oldest of those stars creates activity signals of low amplitude and at much longer periods than the orbital periods of potential planets in the HZ, an ideal case for modern techniques of stellar activity mitigation \citep{Masca2020}. Furthermore, their large planet-star contrast makes them easier targets for atmospheric characterisation \citep{Lovis2017}. Lastly, their large number compared to other spectral types means that there are enough potentially habitable planets located at large angular separations. These large angular separations allow for atmospheric characterisation using high-contrast imaging coupled with high-resolution spectroscopy with next generation giant telescopes, such as the  Extremely Large Telescope (ELT; \citealt{Gilmozzi2007}) or the Thirty Meter Telescope (TMT; \citealt{Sanders2013}) \citep{Lovis2017}, or using nulling interferometry with space missions such as Large Interferometer For Exoplanets (LIFE) \citep{Quanz2022}. 

Here we report the discovery of two Earth-mass planets orbiting the nearby M5.5V star GJ~1002 \citep{Walker1983}, located at just 4.84 pc from the Sun \citep{GaiaEDR3}. The two planets are located within the HZ of the star. Both planets  orbit at angular separations large enough to make them potential candidates for atmospheric characterisation via HDC spectroscopy with ANDES at the ELT \citep{Marconi2021}.

\section{Observations \& data}

GJ~1002 is part of the ESPRESSO and CARMENES GTO programmes \citep{Hojjatpanah2019, Quirrenbach2014}. It has been intensively monitored from 2017 to 2019 by the CARMENES consortium, and independently from 2019 to 2021 by the ESPRESSO consortium, with some months of overlap. 

The ESPRESSO consortium is a collaborative effort between Spanish, Swiss, Portuguese and Italian institutions. It is mainly focused on the search and characterisation of exoplanets \citep{GonzalezHernandez2018, Ehrenreich2020, Pepe2021}, while dedicating a fraction of its observing time to the measurement of fundamental constants of the universe \citep{Schmidt2021, Murphy2022}. The CARMENES consortium is a collaboration of 11 German and Spanish institutions. It is focused on the search and characterisation of exoplanets orbiting M-type stars \citep{Quirrenbach2014, Reiners2018, Zechmeister2019}. 

\subsection{ESPRESSO}

The Echelle SPectrograph for Rocky Exoplanets and Stable Spectroscopic Observations (ESPRESSO) \citep{Pepe2021} is a fibre-fed high-resolution \'Echelle spectrograph installed at the 8.2m ESO Very Large Telescope array, at the Paranal Observatory (Chile). It has a resolving power of R$\sim$140 000 over a spectral range from $\sim$380 to $\sim$780 nm and has been designed to attain extremely high long-term radial-velocity precision. It is contained in a temperature- and pressure-controlled vacuum vessel to avoid spectral drifts due to temperature and air pressure variations. Observations are carried out with simultaneous calibration, using a Fabry-P\'erot etalon. The use of the FP enables correcting any remaining instrumental drift up to a precision of 10 cm~s$^{-1}$ \citep{Wildi2010}. ESPRESSO is equipped with its own pipeline, providing extracted and wavelength-calibrated spectra, as well as RV measurements, other data products such as cross-correlation functions (CCF) and telemetry data. 

We obtained 53 ESPRESSO observations between January 2019 and December 2021. The observations were obtained with a typical exposure time of 900s. In June 2019, ESPRESSO underwent an intervention to update the fibre link, improving the instrument’s efficiency by up to 50\% \citep{Pepe2021}. This intervention introduced an RV offset, leading us to consider separate ESPRESSO18 and ESPRESSO19 datasets. More recently, operations at Paranal were interrupted due to the COVID-19 pandemic and ESPRESSO was taken out of operations between 24 March 2020 and 24 December 2020. This led to a large gap in the observations after the initial campaign. Moreover, a change in one of the calibration lamps after the ramp-up of the instrument is likely to have introduced another RV offset. Therefore, we consider an independent ESPRESSO21 dataset for data obtained after the ramp-up. In summary, we have 53 available RVs, divided between ESPRESSO18 (3 points), ESPRESSO19 (12 points) and ESPRESSO21 (38 points) subsets.

\subsection{CARMENES}

The Calar Alto high-Resolution search for M-dwarfs with Exoearths with Near-infrared and optical Échelle Spectrographs (CARMENES) \citep{Quirrenbach2014} consists of visual (VIS) and near-infrared (NIR) vacuum-stabilised spectrographs covering 520 -- 960 nm and 960 -- 1710 nm with a spectral resolution of 94~600 and 80~400, respectively. It is located at the 3.5~m Zeiss telescope at the Centro Astronómico Hispano Alemán (Almería, Spain).  We extracted the spectra with the \texttt{CARACAL} pipeline, based on flat-relative optimal extraction \citep{Zechmeister2014}. The wavelength calibration that was performed by combining hollow cathode (U-Ar, U-Ne, and Th-Ne) and Fabry-P\'erot etalons exposures. The instrument drift during the nights is tracked with the Fabry-P\'erot in the simultaneous calibration fibre. We obtained 86 observations between 2017 and 2019, using only the data obtained with the CARMENES visual arm. A combination of low RV content in the NIR \citep{Reiners2018} and the more complicated NIR RV extraction \citep{Zechmeister2019} makes it difficult to derive high precision RVs using the CARMENES NIR arm. The instrumental limit of the CARMENES NIR arm in the years 2017 to 2019 was $\sim$ 3.7 m~s$^{-1}$ \citep{Bauer2020}, which makes it unfeasible using them to study the presence of $\sim$ 1 m~s$^{-1}$ signals.

\subsection{Radial velocities}

The radial velocity measurements were obtained using the \texttt{SERVAL} algorithm \citep{Zechmeister2018} for both instruments. This software builds a high signal-to-noise template by co-adding all the existing observations, and performs a least-squares minimisation of each observed spectrum against the template, yielding a measure of the Doppler shift and its uncertainty. We obtained typical RV internal errors of 0.3 and 1.5 m~s$^{-1}$ for ESPRESSO and CARMENES VIS measurements, respectively. We measured an RMS of the RVs of 1.7 m~s$^{-1}$ and 2.5 m~s$^{-1}$ for the ESPRESSO and CARMENES VIS, respectively. Figure~\ref{gj1002_data} shows the RV time series obtained for GJ~1002. 

\subsection{Full-width half maximum of the CCF}

Stellar activity is one of the most important sources of false positive detections when searching for the presence of exoplanets in RV time series. In most stars, the presence of long-lived large spots (or spot groups) on the stellar surface often creates periodic signals in the data that can easily be mistaken for planetary signals \citep{QuelozHenry2001, Robertson2014}. 

\begin{figure}
    \includegraphics[width=9cm]{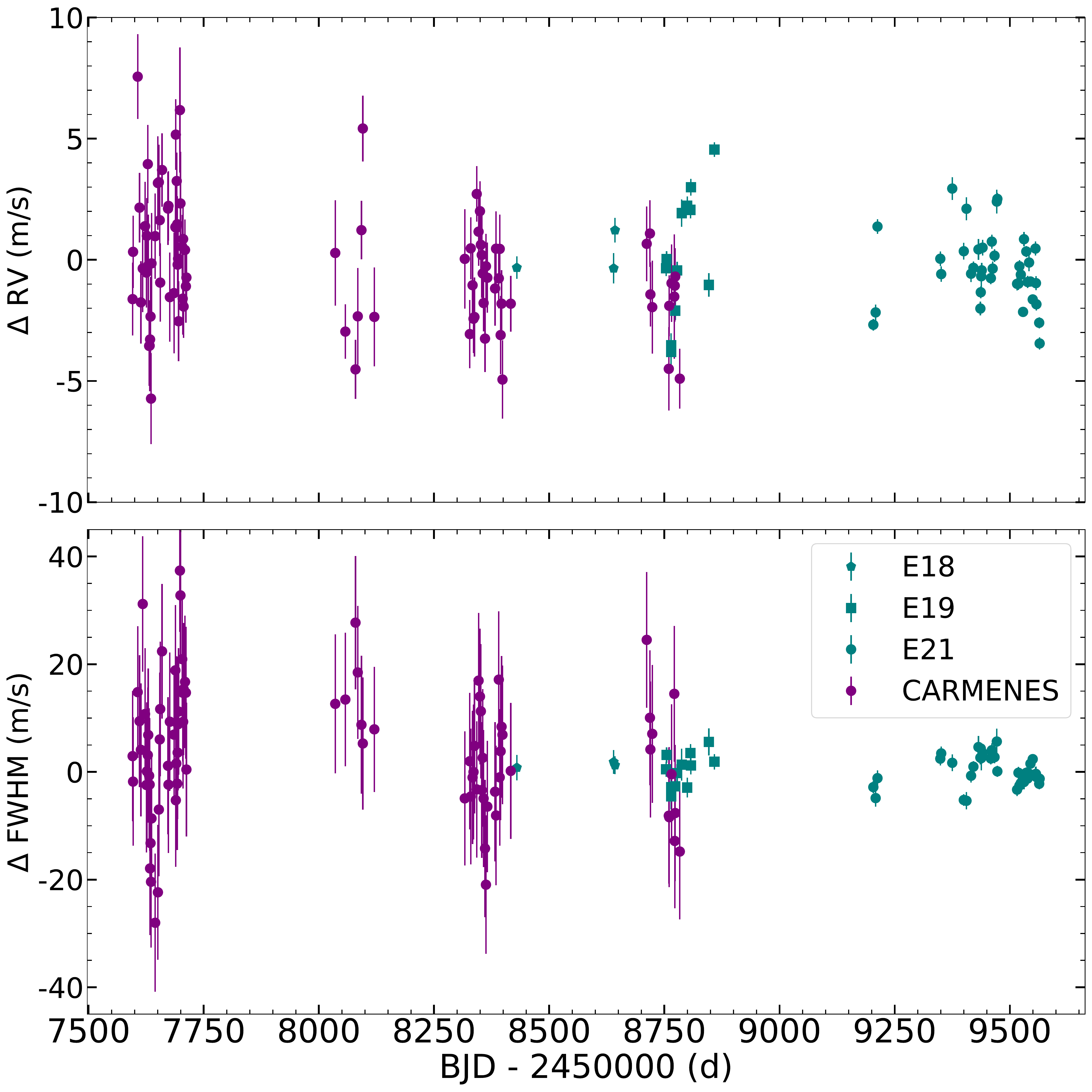}
	\includegraphics[width=9cm]{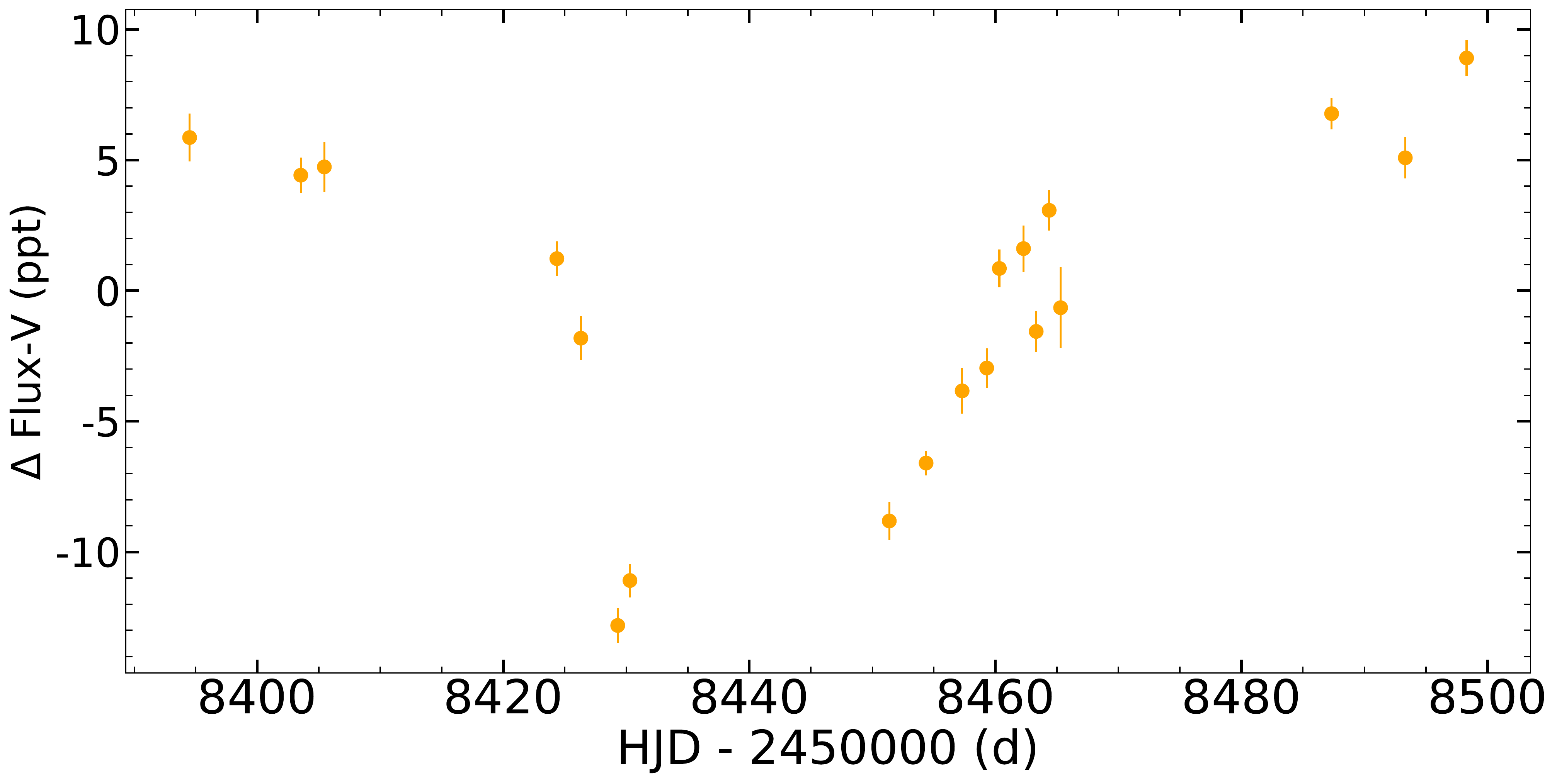}
	\caption{Scientific data used in the subsequent analysis. \textit{Top panel}: RV data of GJ~1002. The purple filled circles show the CARMENES data. The teal symbols show the ESPRESSO data. \textit{Middle panel}: FWHM time series of GJ~1002. \textit{Bottom panel}: SNO V-band photometry of GJ~1002.}
	\label{gj1002_data}
\end{figure}

To monitor the star’s behaviour, we constructed time series of the full-width half maximum (FWHM) of the CCF of the spectra with a binary mask computed from a stellar template \citep{Baranne1996, PepeMayor2000}. The mask was created using an ESPRESSO spectrum of Proxima  as template (same spectral type as GJ~1002, but much higher signal-to-noise). Lines were identified through an automatic line-searching algorithm based on the spectrum derivative. The FWHM of the CCF has been successfully used to model the activity of stars of similar spectral type (e.g. \citealt{Masca2020,Lillo-Box2020} or \citealt{Faria2022}). In the case of some low-mass stars it accurately tracks brightness changes of the star in a similar way as high precision photometry does \citep{Masca2020}. This change in FWHM is usually related to the RV trough its gradient, same as the photometry \citep{Aigrain2012}. The FWHM of the CCF in ESPRESSO is automatically provided by the ESPRESSO Data Reduction Software (DRS). The FWHM of the CCF in the CARMENES VIS data were calculated from a CCF built using the same binary mask as used by the ESPRESSO DRS. Figure~\ref{gj1002_data} shows the resulting CCF time series. The ESPRESSO data provided a much cleaner FWHM time series, with an RMS of 2.8 m~s$^{-1}$ and a typical uncertainty of 1.4 m~s$^{-1}$, while also showing some structure recognisable by eye. The CARMENES FWHM time series show an RMS of 12 m~s$^{-1}$ with a typical uncertainty of 12 m~s$^{-1}$. 

We studied the behaviour of several other activity indicators, which were not included in the global analysis. Most of them show no significant periodicities, or do not provide information in addition to that contained in the FWHM data. Their description and periodograms can be found in Appendix~\ref{append_activity}.

\subsection{Sierra Nevada Observatory photometry}

As part of the follow-up during the CARMENES observations, a short photometric time series was obtained using the telescopes at the Sierra Nevada Observatory (SNO). The T150 telescope is a 1.5 m Ritchey-Chrétien telescope equipped with a CCD camera VersArray 2k×2k, with a field of view of 7.9×7.9 arcmin$^{2}$ \citep{RodriguezSNO2010}. A set of observations was collected in Johnson $V$, $R$, and $I$ filters, consisting in of 19 epochs obtained during the period July 2017 to August 2017. While the baseline is very short, the cadence is dense enough to show clear variations that might be related to one rotation. Each epoch is an average of $\sim$ 19 individual sub-exposures. Figure~\ref{gj1002_data} (bottom panel) shows the V-band photometric time series. The V-band data show the highest dispersion among all filters, with an RMS of 6 ppt and a typical precision of 0.8 ppt.
http://cdsweb.u-strasbg.fr/cgi-bin/qcat?J/A+A/

\subsection{TESS photometry}

GJ~1002 (TIC 176287658) was observed by the Transiting Exoplanet Survey Satellite (\textit{TESS}) \citep{Ricker2015} with the 2\,min cadence in sector 42. The data were processed by the Science Processing Operations centre (SPOC) pipeline \citep{Jenkins2016} and searched for transiting planet signatures with an adaptive, wavelet-based transit detection algorithm \citep{Jenkins2002, Jenkins2010}.

\begin{figure}
	\includegraphics[width=9cm]{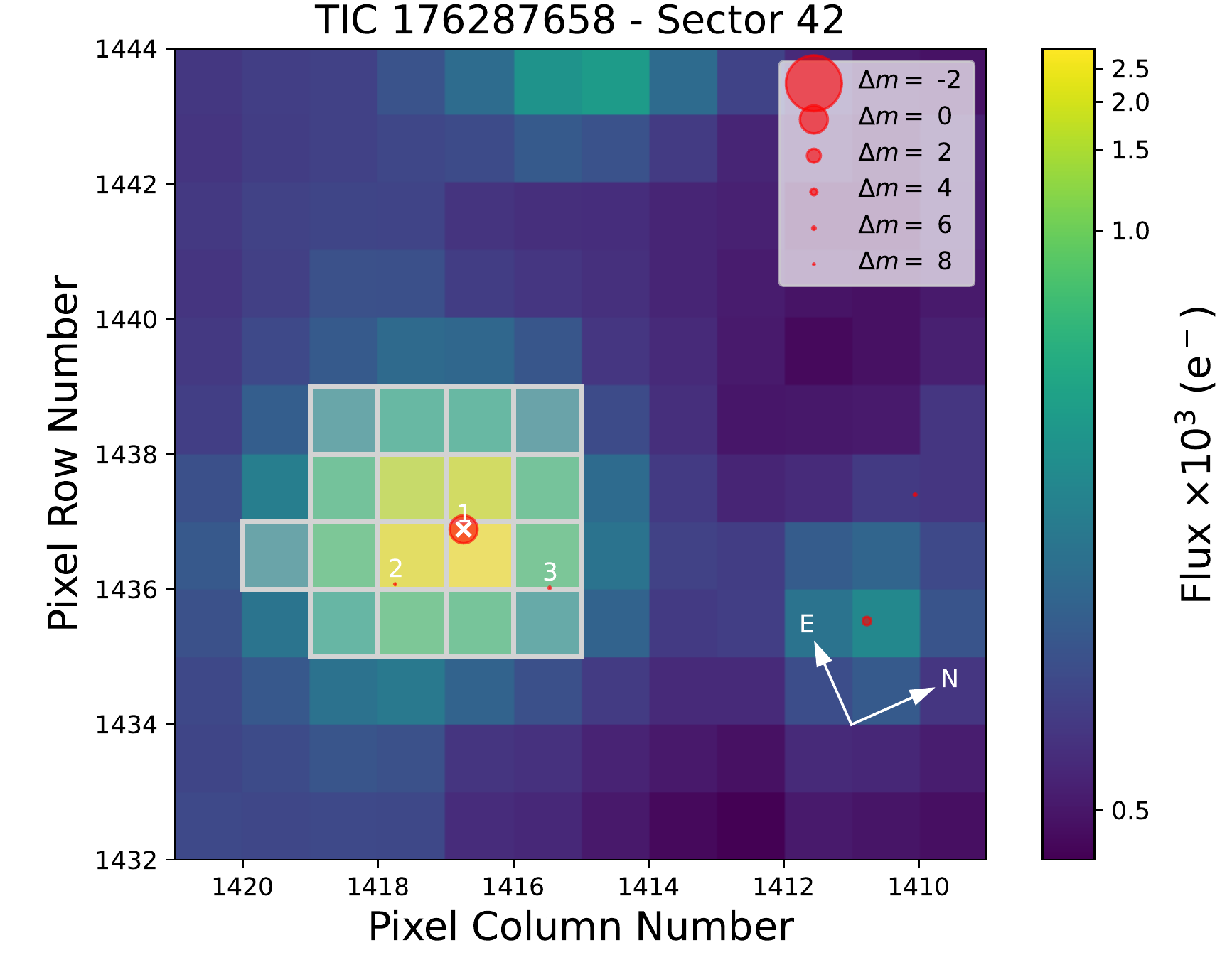}
     \includegraphics[width=8.5cm]{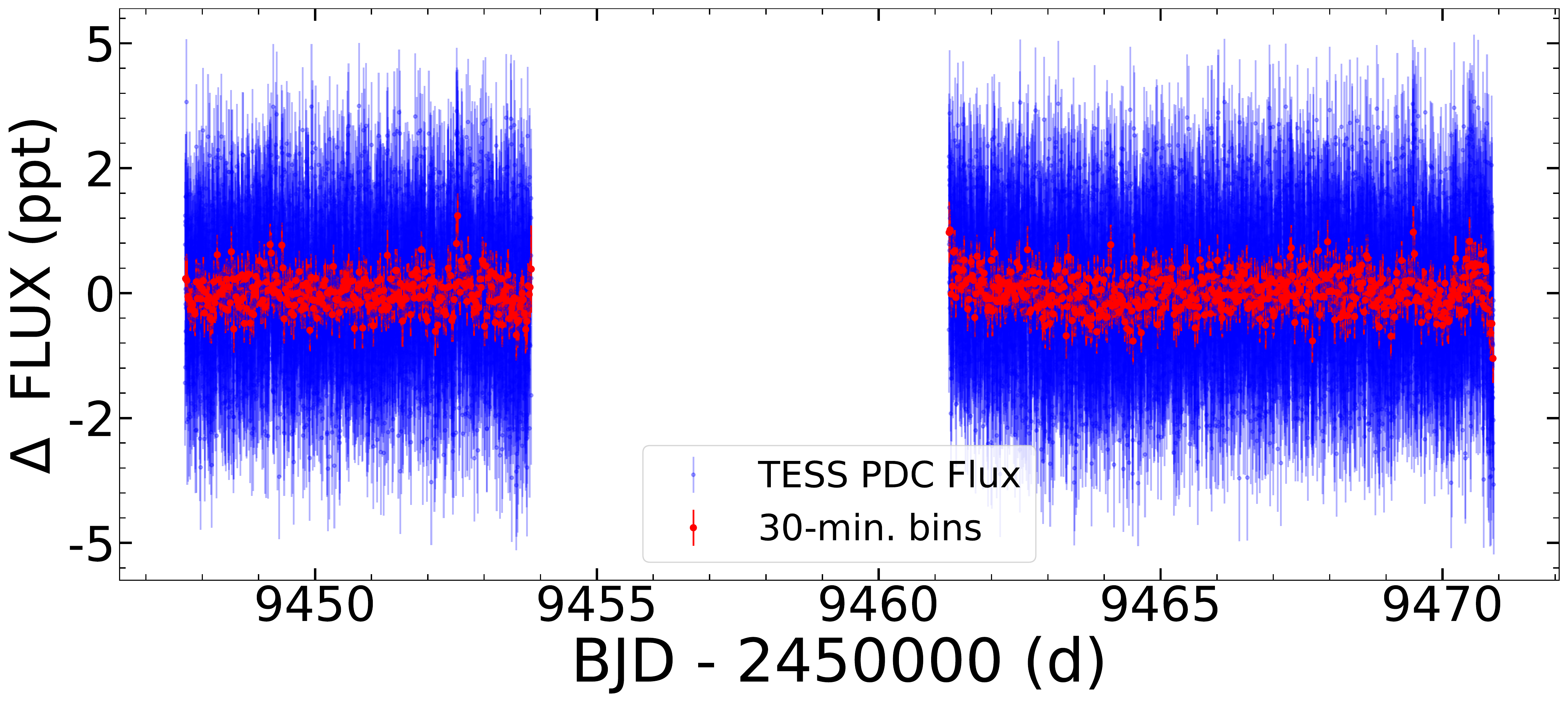}
	\caption{\textit{TESS} data. \textit{Top panel}: TPF file of GJ1002 (cross symbol) for \textit{TESS} sector 42. The electron counts are colour-coded. The \textit{TESS} optimal photometric aperture used to obtain the SAP fluxes is marked with white squares. The \textit{Gaia} DR2 objects with $G$-band magnitudes down to 8\,mag fainter than GJ1002 are labelled with numbers, and their scaled brightness, based on \textit{Gaia} magnitudes, is shown by red circles of different sizes (see figure inset). The pixel scale is 21\,arcsec\,pixel$^{-1}$. \textit{Bottom panel}: TESS light curve. Blue symbols show the 2-minute cadence data. The red symbols represent 30-minute bins.}

	\label{tess_data}
\end{figure}

We verified that the \textit{TESS} fluxes automatically computed by the pipeline are useful for scientific studies by confirming that no additional bright source contaminates the aperture photometry. Figure~\ref{tess_data} displays the target pixel file (TPF) of GJ~1002 using the publicly available \texttt{tpfplotter} code \citep{Aller2020}. This code overplots all sources from the \textit{Gaia} Data Release 3 (DR3) catalogue \citep{GaiaEDR3} with a magnitude contrast up to $\Delta m$=8\,mag on top of the \textit{TESS} TPFs. There are two very faint ($\Delta m$=8\,mag) \textit{Gaia} sources within the photometric aperture around GJ~1002 automatically selected by the pipeline. Therefore, we considered the extracted \textit{TESS} light curve to be free of significant contamination from nearby stars.  The TESS 2-minute cadence data showed a dispersion of 1.25 parts per thousand (ppt), with a median uncertainty per observation of 1.26 ppt. The binned light curve, with 30-minute cadence, showed a dispersion of 0.37 ppt with a typical uncertainty of 0.45 ppt.  

We applied the box least squares (BLS) periodogram \citep{Kovacs2002, Hartman2016} to the \textit{TESS} time series data to search for transit features. No transits have been found in the BLS periodogram. We did not observe any significant stellar variability in the \textit{TESS} light curve, or the presence of flares.

\section{Stellar characteristics} \label{stellar_params}

GJ~1002 is a faint ($V\sim$13.8~mag) nearby M5.5V star located at 4.84 pc from the Sun (calculated using the parallax from \citet{GaiaEDR3}). It has a mass of 0.12 M$_{\odot}$ and colours $B-V$ $\sim$ 1.8, and $V-J$ $\sim$ 5.5, very similar to Proxima. Following \citet{Masca2015}, we measure a log$_{10}$ R$_{HK}^{'}$ = --5.7 $\pm$ 0.2, which corresponds to a rotation period of $\sim$ 130 days \citep{Masca2018}, suggesting that it is older than Proxima. Another characteristic pointing to its old age is its X-ray emission (log $L_{X}$ < 25.54). GJ~1002 is the star with the lowest $L_{X}$ in the NEXXUS survey \citep{SchmittLiefke2004}, pointing once more to an old age.
Table ~\ref{tab:parameters} shows the full set of parameters of GJ~1002. We calculated the mass by using the effective temperature ($T_{\rm eff}$; \citealt{Passegger2022}) together with the luminosity \citep{Cifuentes2020} by following \citet{Schweitzer2019}.

Using the stellar mass and luminosity, we estimate the optimistic HZ for moist worlds to extend from 0.033 to 0.82 au \citep{Kopparapu2014, Kopparapu2017}, which corresponds to orbital periods of 6.6 to 24.8 days. The inner edge of the HZ for worlds with very little water content could extend inwards to 0.023 au \citep{Zsom2013}, or 0.017 au in the case of high albedo, which corresponds to orbital periods of 3.7 and 2.33 days, respectively. 

\begin{table}
\begin{center}
\caption{Stellar properties of GJ~1002 \label{tab:parameters}}
\begin{tabular}[centre]{l l l}
\hline \hline
Parameter & GJ~1002 & Ref. \\ \hline
$\alpha$ (J2000) & 	00:06:43.197 & 1 \\
$\delta$ (J2000) & --07:32:17.019 & 1\\
$\mu_{\alpha}$ (mas yr$^{-1}$)& --811.566 & 1 \\
$\mu_{\delta}$ cos $\alpha$ (mas yr$^{-1}$)& --1893.251 & 1 \\
$\varpi$ ($mas$) &  	206.350 $\pm$ 0.047 & 1\\
d (pc) & 4.8461 $\pm$ 0.0011 & 0 \\
$V$	 (mag) & 13.837 $\pm$ 0.003 & 2 \\
$J$	 (mag) & 8.323 $\pm$ 0.019 & 3 \\
Spectral Type  & M5.5~V & 4\\
$L_{*}$ ($L_{\odot}$) & 0.001406 $\pm$ 0.000019 & 5 \\
$T_{\rm eff}$ (K) & 3024 $\pm$ 52 & 6 \\
$[\rm Fe/H]$ (dex) & -0.25 $\pm$ 0.19 & 6 \\
log $g$ (cgs) & 5.10 $\pm$ 0.06 & 6\\
$R_{*}$ (R$_{\odot}$) & 0.137  $\pm$ 0.005 & 0 \\
$M_{*}$ (M$_{\odot}$) & 0.120  $\pm$ 0.010 & 0 \\
log $L_{X}$ (erg~s$^{-1}$)  & < 25.54  & 8 \\
log$_{10}$ R$_{HK}^{'}$ & -- 5.7 $\pm$ 0.2 & 0 \\
$P_{rot}$ (days) & 126 $\pm$ 15 & 0$^{*}$ \\
\hline
\end{tabular}
\end{center}
\textbf{References:} 0 - This work, 1 -  \citet{GaiaEDR3}, 2 - \citet{Zacharias2013}, 3 - \citet{Cutri2003}, 4 - \citet{Walker1983}, 5 - \citet{Cifuentes2020}, 6 -  \citet{Passegger2022}, 7 - \citet{Schweitzer2019}, 8 - \citet{SchmittLiefke2004}, $^{*}$Rotation period obtained from the global analysis of the FWHM and RV data. 
\end{table}

\section{Analysis} 

Using the data described above, we constructed a dataset that consists of radial velocities and several activity proxies. Figure~\ref{gj1002_data} shows the data that will later be used for the modelling after selecting the most favourable activity indicator (FWHM). Appendix~\ref{append_activity} shows the data of the other activity indicators. 

\subsection{Telemetry data}

Modern RV spectrographs are designed to minimise instrumental effects caused by the changes in their environments. However, we attempt to measure very low amplitude radial velocity signals. Small effects either linked to the stability of both instruments or the extraction of the velocities (e.g trends with airmass) can be present in the data, potentially biasing some of the results. We studied possible correlations between the measured quantities (RV, activity indicators) and all the telemetry data stored in the headers of the reduced ESPRESSO and CARMENES spectra, as well as the periodicities present in the time series of the telemetry data. As both instruments save different data, we performed the analysis independently on both datasets. 

\begin{figure}[ht]
	\includegraphics[width=9cm]{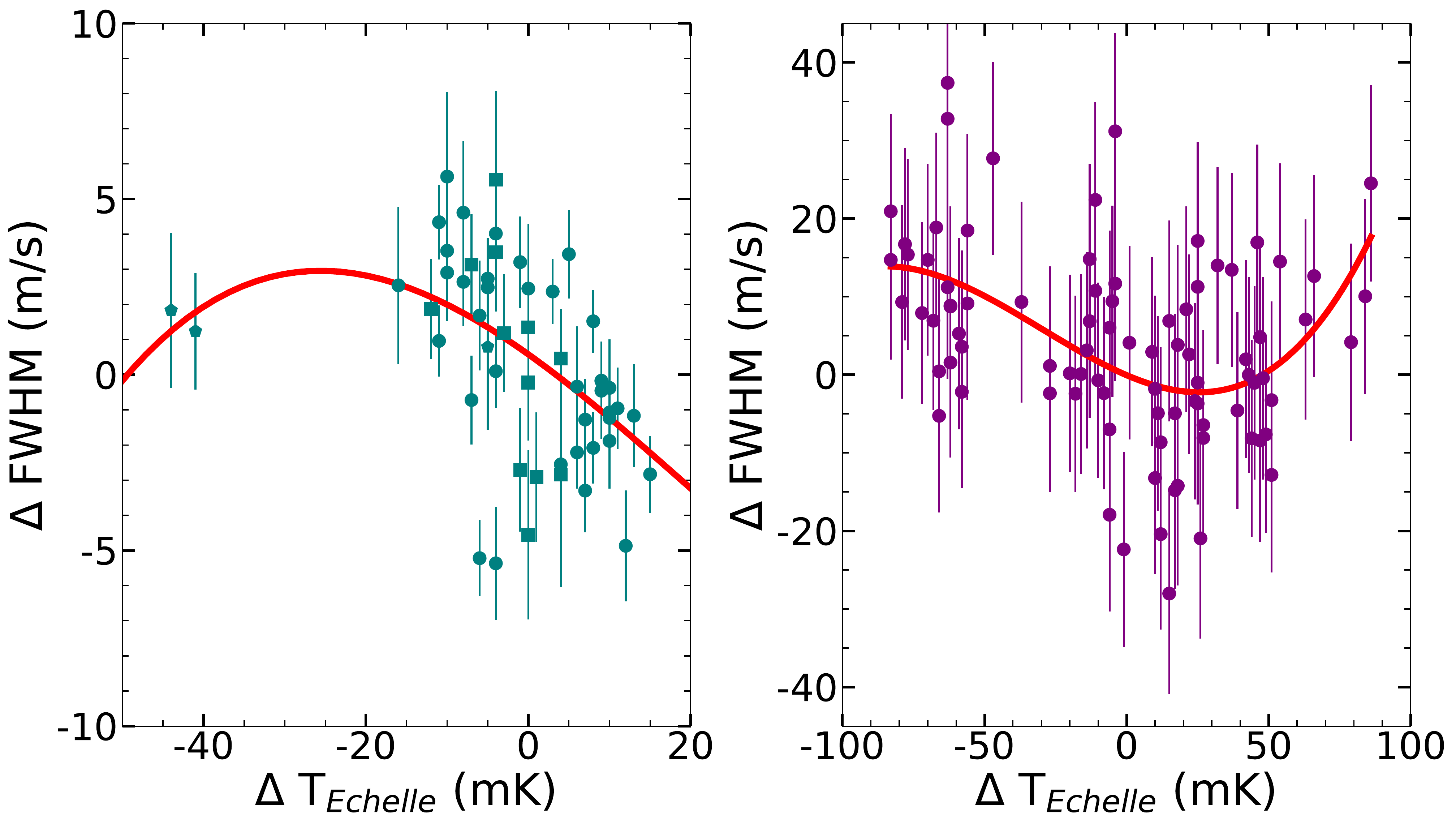}
 \caption{FWHM data of ESPRESSO (left) and CARMENES (right) against temperature change of the \'Echelle grating of each instrument.}	
	\label{fwhm_echelle}
\end{figure}

\begin{figure}[ht]
	\includegraphics[width=9cm]{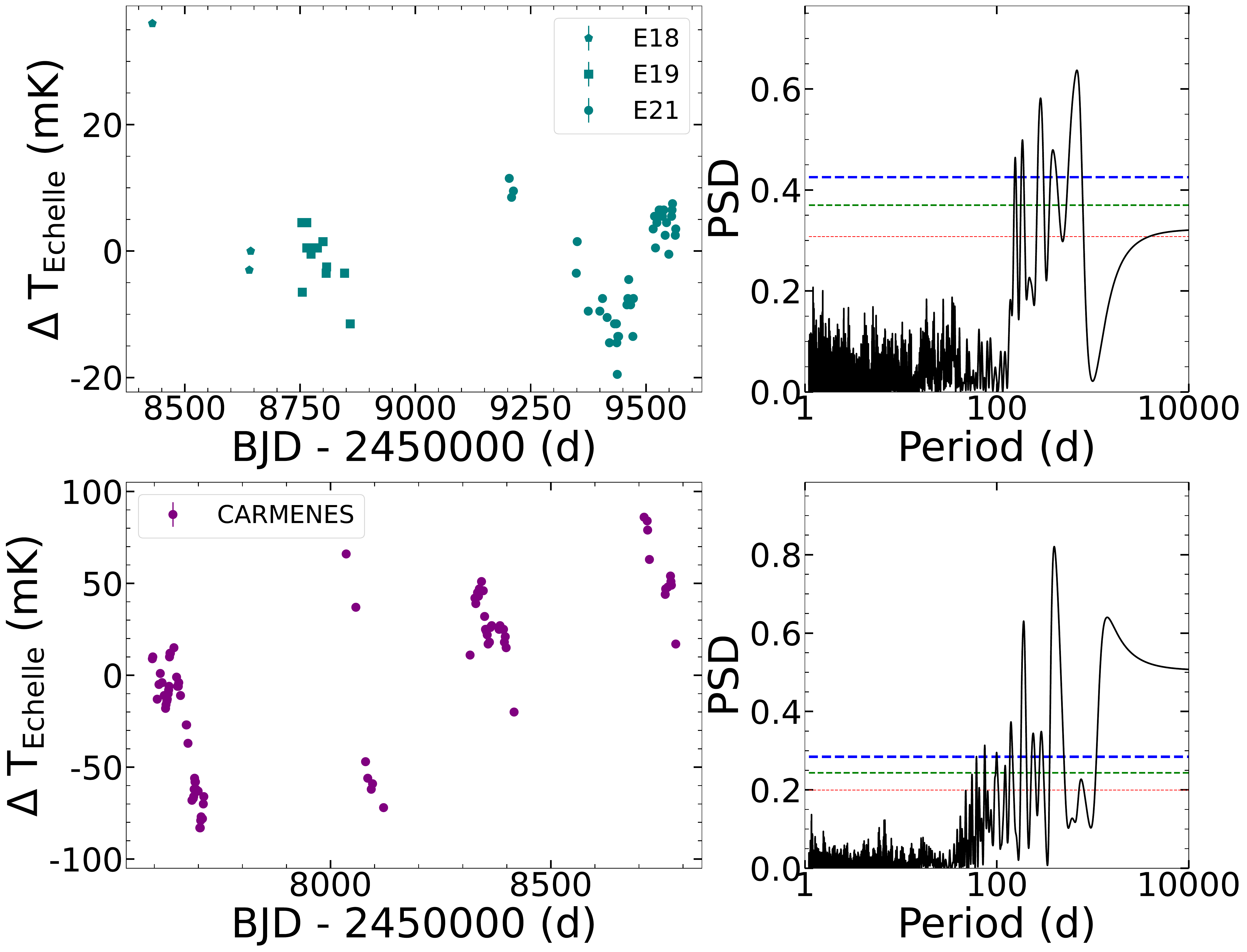}
 \caption{Temperature of the \'Echelle gratings of ESPRESSO (top) and CARMENES (bottom), with their periodogram.}	
	\label{temp_echelle}
\end{figure}

\begin{figure*}[ht]
	\includegraphics[width=18cm]{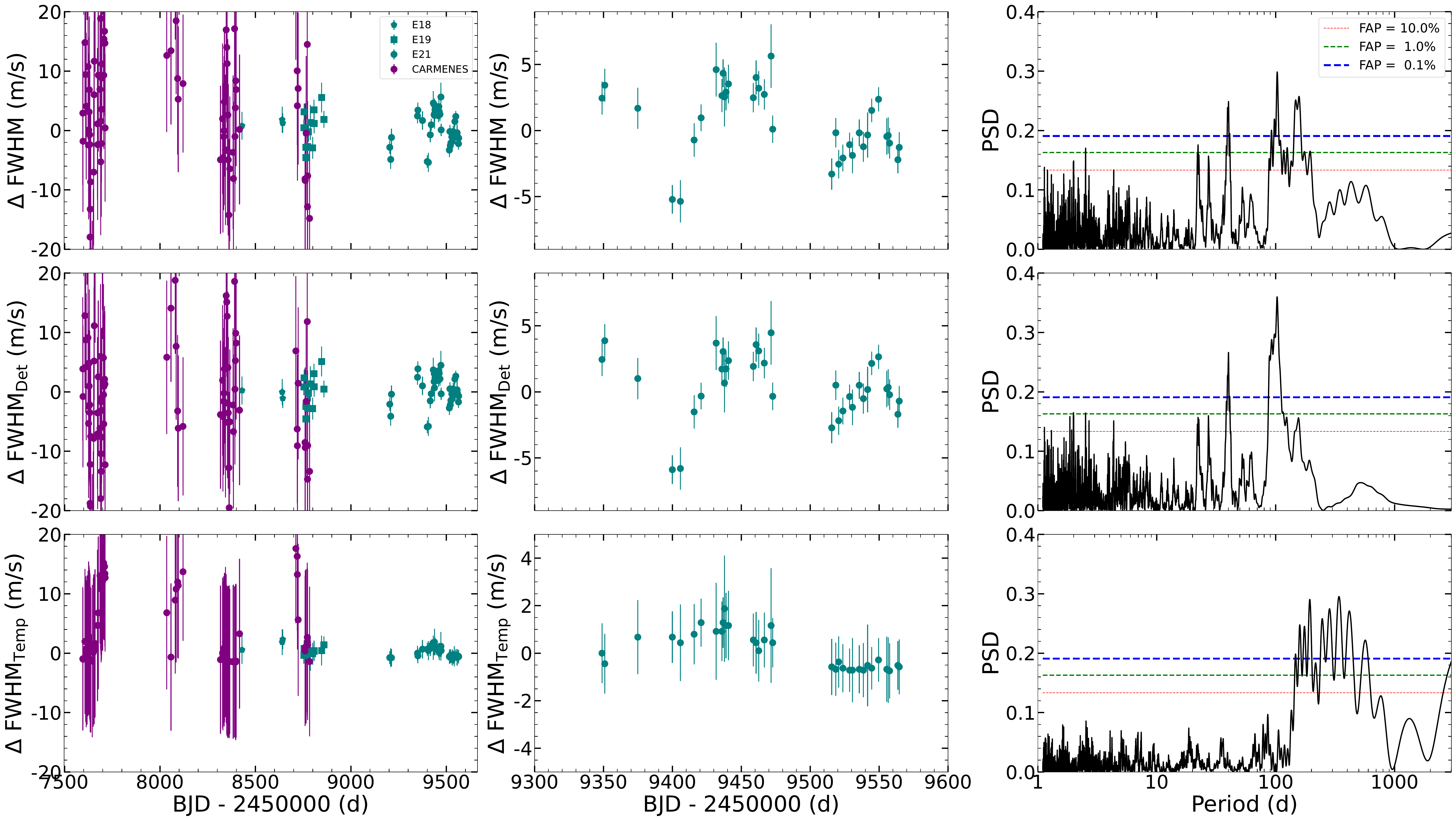}
	\caption{Effect of the changes of the temperature of the \'Echelle grating in the FWHM. \textit{Top panels}: Combined FWHM time series of ESPRESSO and CARMENES, a zoom into the latest ESPRESSO campaign and the periodogram of the data. \textit{Middle panels}: Same plot for the combined FWHM time series of ESPRESSO and CARMENES after detrending with respect to the temperature variations of the \'Echelle grating. \textit{Bottom panels}: Same plot for the estimated effect in FWHM due to temperature changes.}
	\label{fwhm_det}
\end{figure*}

We find a correlation between the FWHM of the CCFs and the temperature of some of the optical elements. Figure~\ref{fwhm_echelle} shows the FWHM measurements compared to the variations in the temperature of the \'Echelle gratings, with respect to the median value for each instrument. We selected this specific measurement as it is one of the common sensors between the two spectrographs in which the effect is present. We used a third-order polynomial to model the correlation between the variations in FWHM and in temperature, as it gives the minimum $\chi^{2}_{\nu}$ of the tested polynomial models. Figure~\ref{temp_echelle} shows the variations of temperature in the \'Echelle grating for ESPRESSO and CARMENES, and their Generalised Lomb Scargle (GLS; \citealt{Zechmeister2009}) periodogram. The data show a periodicity of $\sim$ 1 year, which is also present in the individual datasets. The temperature of the ESPRESSO \'Echelle grating shows an RMS of 11 mK across the whole dataset, reduced to 8 mK after the first intervention in 2019. The temperature of the CARMENES \'Echelle grating shows an RMS of 45 mK. The higher precision of the FWHM measurements in the ESPRESSO data makes it easier to see its effect in the measurements. The effect of detrending against this quantity appear clearly in the combined dataset. Figure~\ref{fwhm_det} shows the FWHM data, and their periodograms, before and after detrending them using the third order polynomials shown in Fig.~\ref{fwhm_echelle}, and the estimated effect in FWHM created by the temperature changes. The raw measurements show power at periods of around 400-600 days and half a year. The detrended data show a much cleaner periodogram with a peak at 102 days, with some peaks at one half and one third of that period, which is consistent with the predicted period derived from rotation-activity relationships. We estimate that the temperature variations introduce an RMS of 0.78 m~s$^{-1}$ in the ESPRESSO FWHM data and of 6 m~s$^{-1}$ in the CARMENES FWHM data, with periodicities around one year and its second harmonic (See. Fig.~\ref{fwhm_det}, bottom panels). 

We see similar effects in some other activity indicators, and with respect to the telemetry obtained trough other temperature sensors. We opted for using a third order polynomial relationship between the affected activity indicators and the \'Echelle grating temperature in all subsequent analyses. The parameters of the polynomial are always fitted simultaneously with the rest of the parameters of the model, using normal priors for the polynomial parameters, centred around zero and with a sigma equal to or larger than the peak-to-peak variation over the baseline to the power of $i$, for each order $i$ of the polynomial. The models computed in sections~\ref{analysis_stellar} and~\ref{analysis_planets} provide slopes different from zero for the dependence between FWHM and temperature. 

We find no significant correlations between the RV and the telemetry data. There is some apparent, but not significant, slope between the RV and temperature data. We repeated all the analysis of sections~\ref{analysis_stellar} and~\ref{analysis_planets} including a linear term relating the RV and temperature of the \'Echelle grating. We recover null results for all stages of the modelling process. Within the precision of the data and the modelling scheme, we cannot detect RV drifts caused by instrumental temperature drifts for ESPRESSO or CARMENES. These variations, if present in the raw data, are likely successfully corrected by the simultaneous calibration.

\subsection{Stellar activity}\label{analysis_stellar}

\begin{figure*}[ht]
    \includegraphics[width=18cm]{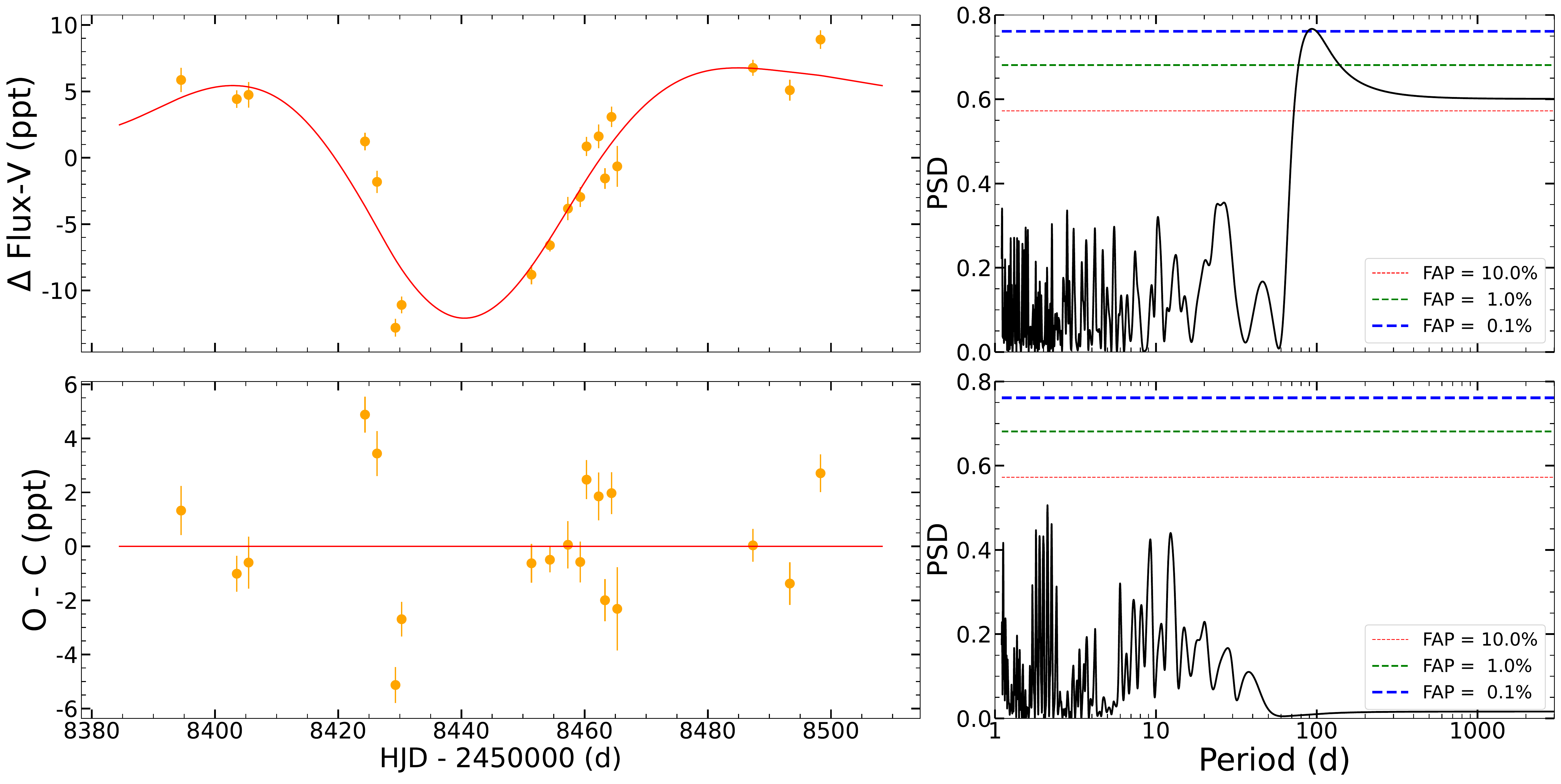}
	\includegraphics[width=18.5cm]{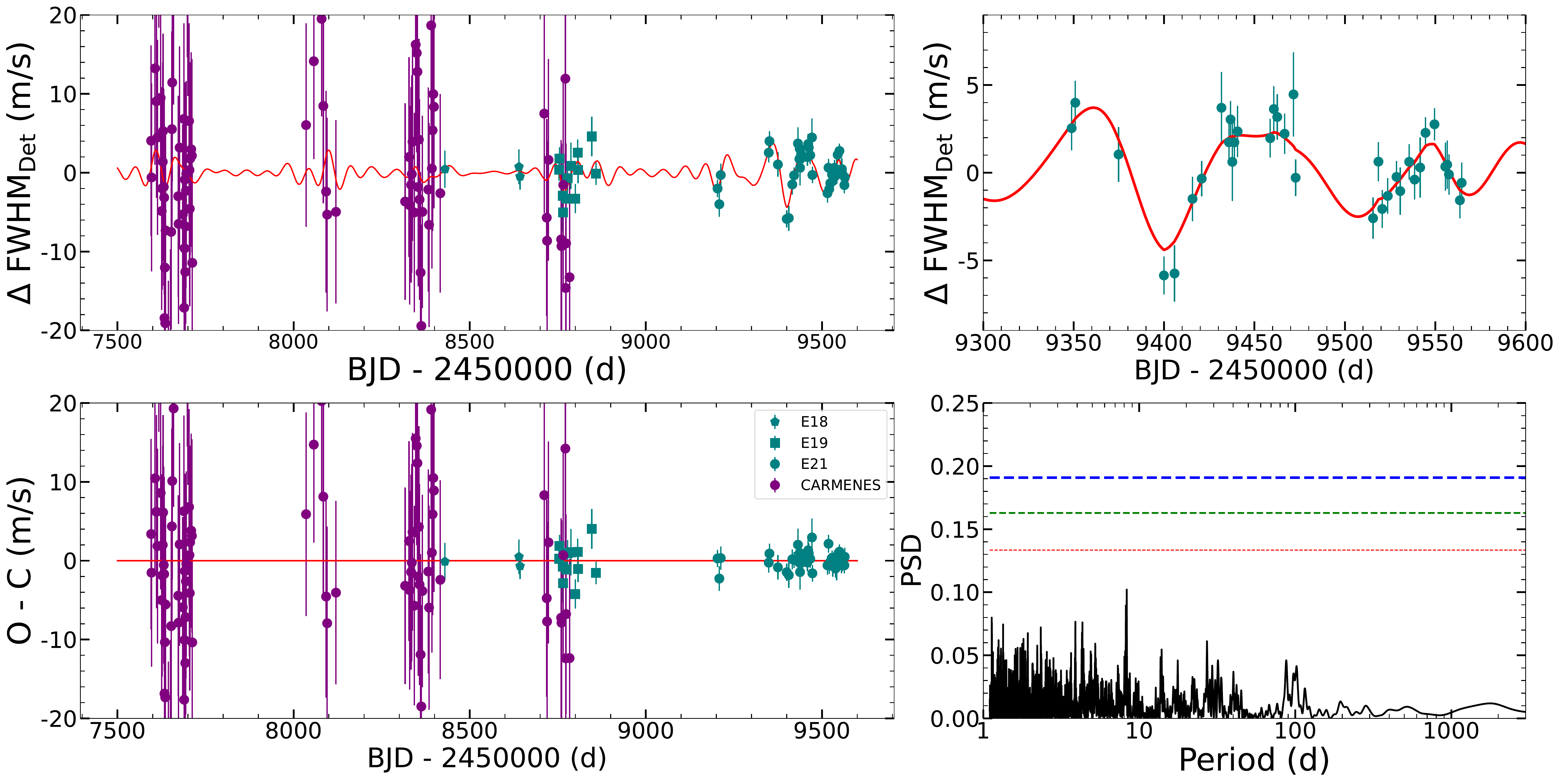}
	\caption{Data for the star GJ~1002, combining ESPRESSO and CARMENES spectroscopy, and SNO V-band photometry. The four upper panels include the V-band photometry data with the best model fit (top left), the periodogram of the data (top right), the residuals after the fit (bottom left) and the periodogram of the residuals. The four lower panels include the FWHM data of ESPRESSO (teal symbols) and CARMENES (purple symbols) with the best model fit (top left), a zoom to the last ESPRESSO campaign (top right), the residuals after the fit (bottom left) and the periodogram of the residuals. The computation of the periodogram uses the offsets and the jitters estimated for model 0-Planets in table~\ref{parameters_1}.}
	\label{sno_fwhm_gp}
\end{figure*}

Stellar activity is one of the biggest causes of false positive detections of exoplanets when using RVs. Therefore we performed the analysis of stellar activity before we search for candidate planetary signals. Based on the $\log_{10}\rm R'_{HK}$ vs P$_{Rot}$ relationship of \citet{Masca2018}, we predict a rotation period of 130 $\pm$ 30 days. The GLS periodogram of FWHM and SNO V-band photometry time series shows an indication of periodicity at a period consistent with the suspected rotation period (Fig.~\ref{fwhm_det} and ~\ref{sno_fwhm_gp}). Using this information, we analysed our activity indicators using the Gaussian processes framework (GP; see \citealt{Rasmussen2006} and \citealt{Roberts2012}). 

The GP framework has become one of the most successful methods in the analysis of stellar activity in RV time series (e.g. \citealt{Haywood2014}). The stellar noise is described by a covariance with a prescribed functional form and the parameters attempt to describe the physical phenomena to be modelled. The GP framework can be used to characterise the activity signal without requiring a detailed knowledge of the distribution of active regions on the stellar surface, their lifetime, or their temperature contrast. One of the biggest advantage of GPs is that they are flexible enough to effortlessly model quasi-periodic signals, and account for changes in the amplitude, phase, or even small changes in the period of the signal. This flexibility is also one of their biggest drawbacks, as they can easily overfit the data, suppressing potential planetary signals. 

There have been significant efforts in recent years to better constrain the GP models to the measurements of the variability seen in the activity indicators. Simple approaches include training them with photometric data or activity indicators (e.g. \citealt{Haywood2014}) or simultaneous modelling of the activity proxies and the radial velocity with shared hyper parameters (e.g. \citealt{Masca2020} or \citealt{Faria2022}). These approaches constrain the covariance matrix used for the RV by using the activity indicators, but still lead to an `independent'  fit of the RV data. A more sophisticated approach is the use of multi-dimensional GPs, which join the fit of all time series under a single covariance matrix (e.g. \citealt{Rajpaul2015}, \citealt{Barragan2022} or \citealt{Delisle2022}), better linking them to each other. The usual assumption is that there exists an underlying function governing the behaviour of the stellar activity, $G(t)$, that manifests in each time series as a linear combination of itself and its gradient, $G'(t)$, with a set of amplitudes for each time series (TS), following the idea of the $FF'$ formalism \citep{Aigrain2012}, as described in equation~\ref{eq_gp_grad}). 

\begin{equation} \label{eq_gp_grad}
\begin{split}
&\Delta ~TS_{1} = A_{1} \cdot G(t) + B_{1} \cdot G'(t)~, \\
&\Delta ~TS_{2} = A_{2} \cdot G(t) + B_{2} \cdot G'(t)~, \\
&...
\end{split}
\end{equation}

\noindent \citet{Masca2020} showed that in the case of Proxima there was a very good correlation between the FWHM of the CCF and the activity-induced RV signal, making these approach very compelling to study the activity signal of other mid- to late-type M-dwarfs, such as GJ~1002.

We used the newly presented \texttt{S+LEAF} code \citep{Delisle2022}, which extends the formalism of semi-separable matrices introduced with \texttt{celerite} \citep{Foreman-Mackey2017} to allow for fast evaluation of GP models even in the case of large datasets. The \texttt{S+LEAF} code allows to fit simultaneously a GP to several time series, based on a linear combination of the GP and its derivative, with different amplitudes for each time series (see equation~\ref{eq_gp_grad}). The \texttt{S+LEAF} code supports a wide variety of GP kernels with fairly different properties. After testing several options we opted for using a combination of two simple harmonic oscillators (SHO) at the rotation period and its second harmonic. See section~\ref{testing} for more details on the alternatives tested. The selected Kernel is defined as: 

\begin{equation} \label{act_model}
\begin{split}
\fontsize{8}{11}\selectfont
 k(\tau) = k_{\rm SHO ~\rm 1}(\tau, P_{1}, S_{1}, Q_{1}) + k_{\rm SHO ~\rm 2}(\tau, P_{2}, S_{2}, Q_{2}) ~,
 \end{split}
\end{equation}

\noindent with $\tau = t - t'$, representing the time-lag between measurements. 

Following equation~\ref{eq_gp_grad}, the activity induced signal in every specific time series is:

\begin{equation} \label{full_gp_model}
\begin{split}
\Delta ~TS = A_{11} \cdot G_{\rm SHO ~\rm 1} + A_{12} \cdot G'_{\rm SHO ~\rm 1} \\
+ A_{21} \cdot G_{\rm SHO ~\rm 2} + A_{22} \cdot G'_{\rm SHO ~\rm 2} ~,
\end{split}
\end{equation}

\noindent where $G_{SHO}$ is the realisation of a GP with kernel $k_{SHO}$ and $G'$ is its first derivative. 

Following \citet{Foreman-Mackey2017}, the $k_{SHO}$ kernel is defined as:

\begin{equation} \label{eq_sho}
\fontsize{8}{11}\selectfont
 k_{i}(\tau) = {C_{i}^{2}} e^{-\tau/L}  \left\{\begin{array}{cc}\cosh(\eta {2 \pi \tau}/P_{i})+{{P_{i}}\over{2 \pi \eta L }}\sinh(\eta {2 \pi \tau}/P_{i}) ; ~\rm if ~P_{i} > 2 \pi L\\2 (1 + {{2 \pi \tau}\over{P_{i}}}) ; ~\rm if ~P_{i} = 2\pi L\\ \cos(\eta {2 \pi \tau}/P_{i}) + {{P_{i}}\over{2 \pi \eta L}} \sin(\eta {2 \pi \tau}/P_{i}); ~\rm if ~P_{i} < 2 \pi L\end{array}\right\}~,
\end{equation}

\noindent with $\eta = (1 - (2L/P_{i})^{-2})^{1/2}$, controlling the damping of the oscillator.

This Kernel has a power spectrum density:  
\begin{equation} \label{psd_kernel}
S(\omega) = \sqrt {{2} \over {\pi}} {{S_{i} ~\omega_{i}^{4}} \over {(\omega^{2} - \omega_{i}^{2})^2 + \omega_{i}^{2}~\omega^{2} / Q^{2}}}~,
\end{equation}

\noindent where $\omega$ is the angular frequency, $\omega_{i}$ is the undamped angular frequency for each component ($\omega_{i}$ = 2 $\pi$ / $P_{i}$), $S_{i}$ is the power at $\omega$ = $\omega_{i}$ for each component, and $Q_{i}$ is the quality factor. $S_{i}$, $P_{i}$ and $Q_{i}$ are the parameters sampled in the covariance matrix, which are related to the amplitude ($C_{i}$), rotation period ($P_{rot}$) and timescale of evolution ($L$) in the following way: 

\begin{equation} \label{eq_params}
\begin{split}
&P_{1} = P_{rot} ~,~ S_{1} = {{C_{1}} \over {2 \cdot L}} \left( {{P_{1}} \over {\pi}} \right)^{2} ~,~  Q_{1} = {\pi {L}\over{P_{1}}}~,~ \\
&P_{2} = {{P_{rot}}\over{2}} ~,~ S_{2} = {{C_{2}} \over {2 \cdot L}} \left( {{P_{2}} \over {\pi}} \right)^{2} ~,~  Q_{2} = {\pi {L}\over{P_{2}}}~.
\end{split}
\end{equation}

\noindent The covariance matrix also includes a term of uncorrelated noise ($\sigma$), independent for every instrument, added in quadrature to the diagonal of the covariance matrix to account for all unmodelled noise components, such as uncorrected activity or instrumental instabilities. 

The amplitudes $C_{i}$ in equations~\ref{eq_sho} and \ref{eq_params} are related to the amplitude of the underlying function, not to any of the specific time series. These amplitudes are degenerate with the amplitudes of each time series. We opted to fix S$_{1}$ and S$_{2}$ to 1, which fixes their power at $\omega$ = 0. The amplitudes of every component will be governed by the parameters A$_{ij}$ shown in equation~\ref{full_gp_model}. 

We performed the analysis of the FWHM and SNO V-band photometry using the suspected rotation period as prior information. \citet{Giles2017} showed that the timescale of evolution of signals in Kepler stars is typically between one and two rotations, sometimes longer for M-dwarfs. We also include this knowledge as prior information, using a log-scale to allow for a long tail to long timescales of persistence of the signals. We use log-normal priors for the amplitudes and jitter terms, centred on $\sim$ln(RMS) of the data and with a sigma of $\sim$ln(RMS) of the data. When using a GP with a completely free amplitude and jitter parameters, on data that includes multiple signals, there is a large risk of the GP absorbing all variations present in the data. Constraining the parameters in this way ensures a smooth GP model, preventing it from overfitting variations at short timescales without fully excluding any region of the parameter space.

We modelled the data using Bayesian inference via nested sampling \citep{Skilling2004}. This allows for an efficient exploration of large parameter spaces, as well as to obtain the Bayesian evidence of the model (i.e. marginal likelihood, lnZ). We used the code \texttt{Dynesty}~\citep{Speagle2020}. \texttt{Dynesty} employs multi-ellipsoidal decomposition \citep{Feroz2009} to more efficiently sample large prior volumes. We used the default configuration, which uses a random walk or random slice \citep{Handley2015a,Handley2015b} sampling strategy depending on the number of free parameters. We set the number of live points equal to $N_{\rm Par}$~$\cdot$~($N_{\rm Par}$+1)/2, and the number of slices equal to 3~$\cdot$~$N_{\rm Par}$ , with $N_{\rm Par}$ being the number of free parameters of the model. 

We find a good solution for the FWHM and SNO data, with a rotation period of 103$^{+30}_{-9}$ and 128$^{+22}_{-19}$ days, respectively. The measured timescales of evolution are 350$^{+400}_{-190}$ and 185$^{+340}_{-108}$ days, respectively. Both indicators seem to have trouble properly constraining the timescale of evolution. In the case of the FWHM, the low precision of the CARMENES FWHM limits the amount of information that can be extracted. The SNO data do not have the baseline to properly constrain the activity timescales. However, it shows that it has to be longer than $\sim$ 90 days. Figure~\ref{sno_fwhm_gp} shows the best fits to the FWHM and SNO data. The analysis supports a rotation period longer than 100 days. We did not find indications of stellar activity variations at different timescales other than the suspected rotation period or its second harmonic. 

\begin{figure*}[ht]
	\includegraphics[width=18cm]{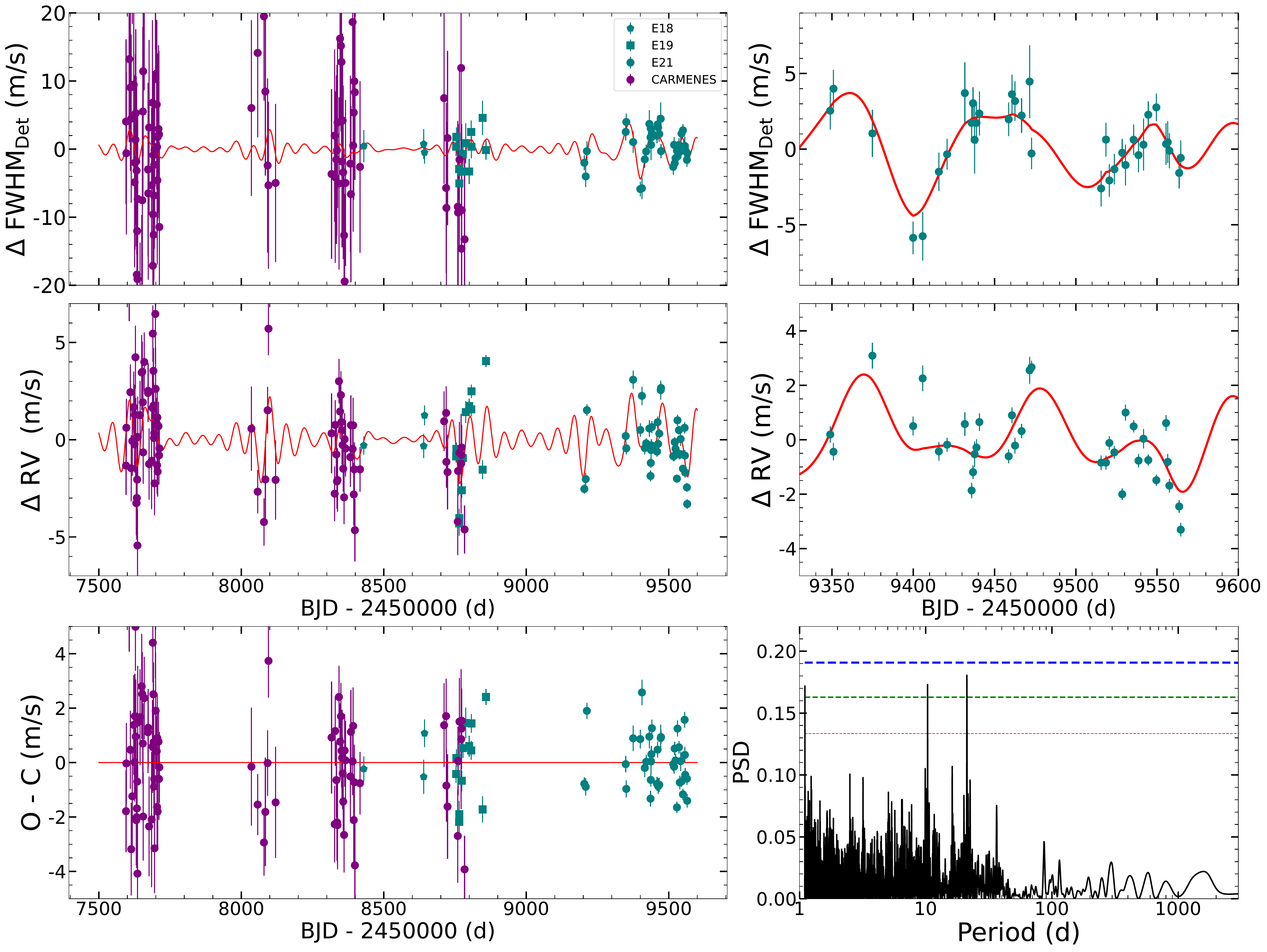}
	\caption{Activity-only model. \textit{Top panels}: Combined FWHM time series of ESPRESSO and CARMENES with the best fit model. The right panel shows a zoom to the latest ESPRESSO campaign. \textit{Middle panels}: Combined RV time series of ESPRESSO and CARMENES with the best fit activity-only model. \textit{Bottom panels}: Residuals after the activity-only fit of the RV time series of ESPRESSO and CARMENES. The right panel shows the GLS periodogram of the residuals. The computation of the periodogram uses the offsets and the jitters estimated for model 0-Planets in table~\ref{parameters_1}.}
	\label{model_rv_0p}
\end{figure*}

We tested a model including FWHM and RV simultaneously, using the same GP model described before. The combined model converges to a rotation period of 120 $\pm$ 12 days with a timescale of evolution of 81$^{+47}_{-32}$ days, consistent with one rotation. The activity parameters are much better constrained after including the RV in the analysis. The models has a measured lnZ of --784.3 $\pm$ 0.4. The residuals after the fit of the RV show an RMS of 1.7 m~s$^{-1}$ (0.9 m~s$^{-1}$ for ESPRESSO), which in principle cannot be accounted for by the uncertainties of the data, and their periodogram shows the presence of 2 periodic signals with false alarm probabilities lower than 1\%, at periods of 10.4 and 21.2 days. As we do not have hints of activity at these timescales, we consider these two signals to be candidate planetary signals. Figure~\ref{model_rv_0p} shows the model in FWHM and RV, the residuals after the fit for the RV, and their GLS periodogram. Table~\ref{parameters_1} shows the priors used and the parameters measured (column labelled as `0 Planets').

\subsection{Candidate planetary signals}\label{analysis_planets}

\begin{figure*}[ht]
	\includegraphics[width=18cm]{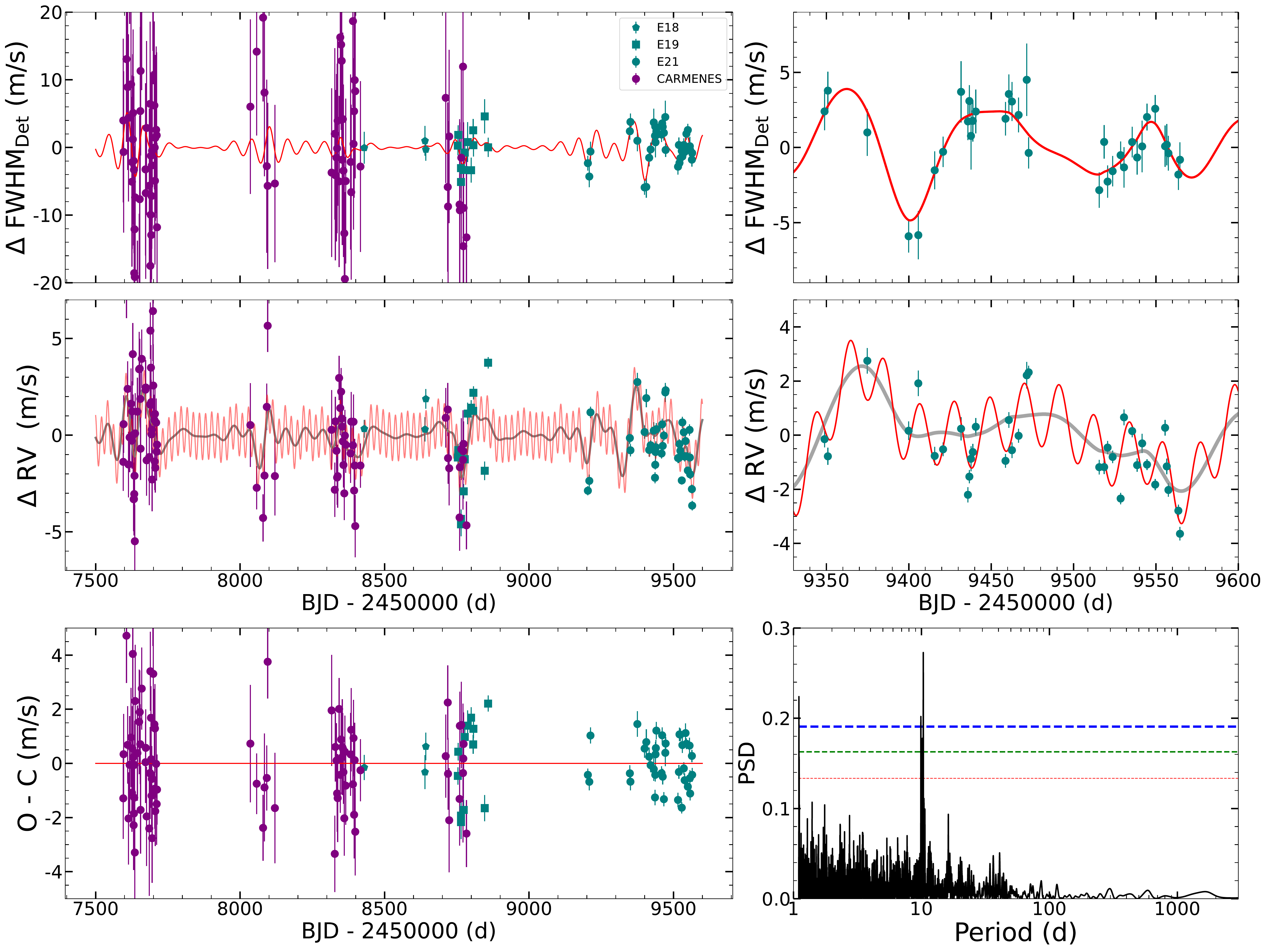}
	\caption{One-planet model. \textit{Top panels}: Combined FWHM time series of ESPRESSO and CARMENES with the best fit model. The right panel shows a zoom to the latest ESPRESSO campaign. \textit{Middle panels}: Combined RV time series of ESPRESSO and CARMENES with the best fit of the GP+1p model. The red line shows the best combined model, while the grey line shows the stellar activity component. \textit{Bottom panels}: Residuals after the fit of the RV time series of ESPRESSO and CARMENES. The right panel shows the GLS periodogram of the residuals. The computation of the periodogram uses the offsets and the jitters estimated for model 1-Planet (Circ) in table~\ref{parameters_1}.}
	\label{model_rv_1p}
\end{figure*}

Following the information given by the periodogram of the residuals in Figure~\ref{model_rv_0p}, we included a sinusoidal signal at a period of 21.2 days. We used a normal prior at 21.2 $\pm$ 4.2 days for the period (20\% of the period for the 1-sigma range), with the time of conjunction parameterised as a phase centred around the mid-point of the observational baseline ($\phi$, $\mathcal{U}$ [0,1]; $T_{0}$ = 8590.096 + $P_{orb} \cdot \phi$), and a uniform prior for the amplitude in the range of 0 -- 10 m~s$^{-1}$. We obtain a fit for a signal with an amplitude of 1.2 $\pm$ 0.2 m~s$^{-1}$. The model has a measured lnZ of --776.9 $\pm$ 0.4. This corresponds to a $\Delta$ lnZ of +7.4 with respect to the 'activity-only' model, which corresponds to a 0.05\% false alarm probability for the more complex model. The activity model in RV changes from the previous iteration, with some of the variations previously attributed to the GP-model being now attributed to the sinusoidal motion. The model of the FWHM does not significantly change by including the sinusoidal signal in the RV. The residuals after the fit in the RV still show an RMS of 1.7 m~s$^{-1}$ (0.9 m~s$^{-1}$ for ESPRESSO), and their periodogram shows a signal at a period of 10.4 days. Figure~\ref{model_rv_1p} shows the best fit to the data along with the residuals and their periodogram. Table~\ref{parameters_1} shows the priors used and the parameters measured (column `1 Planet (Circ)').

We tested for the possibility of the signal being eccentric by including a Keplerian model instead of the sinusoid (eq.~\ref{eq_kepler}).

\begin{equation} \label{eq_kepler}
  y(t)=K \left(\cos(\eta+\omega) + e \ \cos(\omega)\right)
\end{equation} 

\noindent where the true anomaly $\eta$ is related to the solution of the Kepler equation that depends on the orbital period of the planet $P_{\rm orb}$ and the orbital phase $\phi$. This phase corresponds to the periastron time, which depends on the mid-point transit time $T_{0}$, the argument of periastron $\omega$, and the eccentricity of the orbit $e$. We parameterise the eccentricity as $e = (\sqrt{e} ~\cos(\omega))^{2} + (\sqrt{e} ~\sin(\omega))^{2}$ and $\omega = \arctan2(\sqrt{e} ~sin(\omega),\sqrt{e} ~\cos(\omega))$. We then sample $\sqrt{e} ~cos(\omega)$ and $\sqrt{e} ~\sin(\omega)$ with normal priors of 0 $\pm$ 0.3. Eccentricity is typically overestimated in noisy data and datasets with unmodelled sources of variability \citep{Hara2019}. The parametrisation described above favours low eccentricities, expected for the periods of the planets. We measure a lnZ of --777.4 $\pm$ 0.4. This corresponds to a $\Delta$ lnZ of --0.5 with respect to the model using a sinusoidal functions, which favours the circular model. Table~\ref{parameters_1} shows the priors used and the parameters measured (column `1 Planet (Kep)).

\begin{figure*}
	\includegraphics[width=18cm]{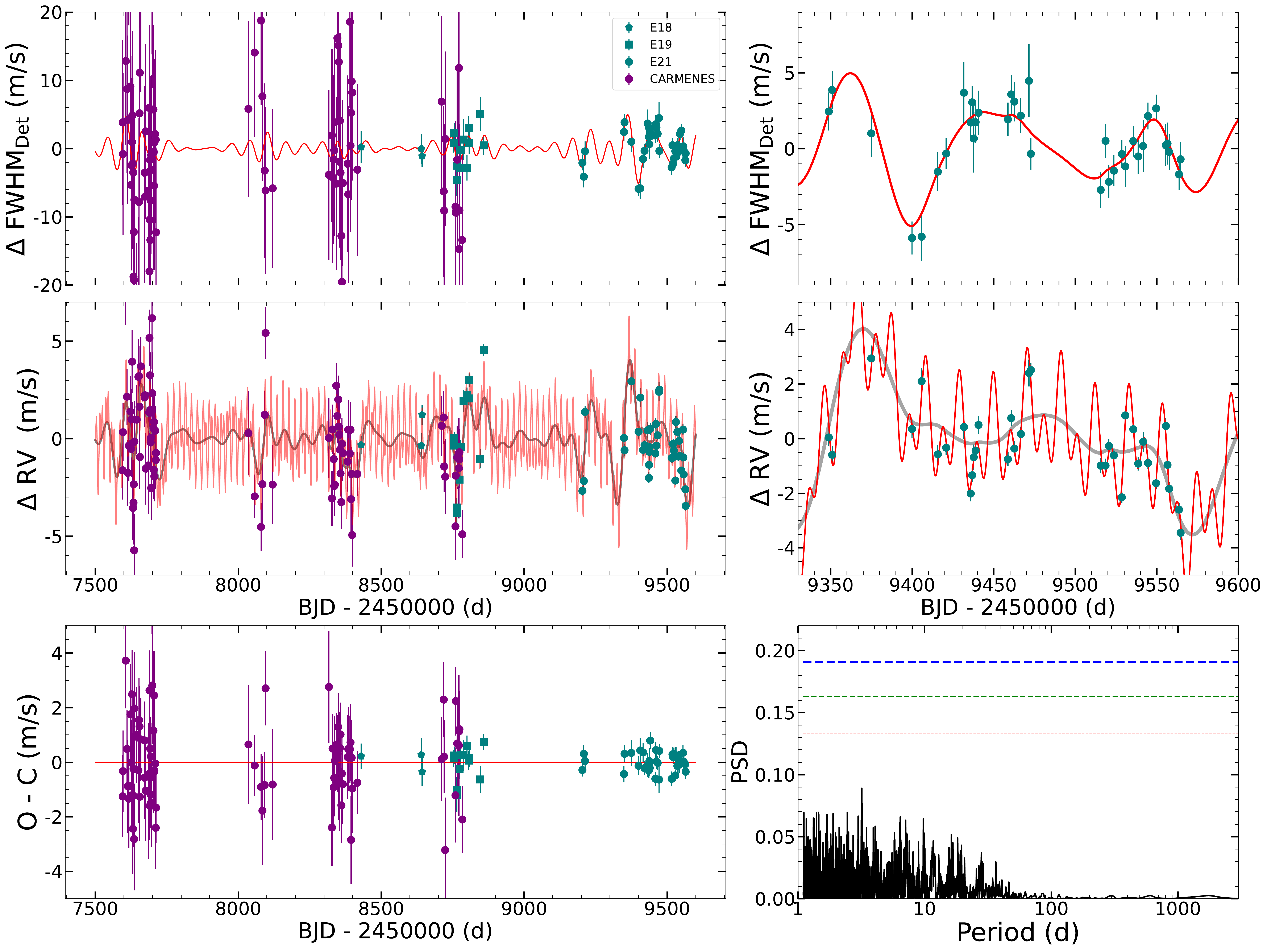}
	\caption{Two-planet model. \textit{Top panels}: Combined FWHM time series of ESPRESSO and CARMENES with the best fit model. The right panel shows a zoom to the latest ESPRESSO campaign. \textit{Middle panels}: Combined RV time series of ESPRESSO and CARMENES with the best fit of the GP+2p model. The red line shows the best combined model, while the grey line shows the stellar activity component. \textit{Bottom panels}: Residuals after the fit of the RV time series of ESPRESSO and CARMENES. The right panel shows the GLS periodogram of the residuals. The computation of the periodogram uses the offsets and the jitters estimated for model 2-Planets (Circ) in table~\ref{parameters_1}.}
	\label{model_rv_2p}
\end{figure*}

We included a second sinusoidal signal using a normal prior at 10.4 $\pm$ 2.1 days for the period, with the time of conjunction parameterised as a phase centred around the mid-point of the observational baseline ($\phi$, $\mathcal{U}$ [0,1]; $T_{0}$ = 8590.096 + $P_{orb} \cdot \phi$), and a uniform prior for the amplitude in the range of 0 -- 10 m~s$^{-1}$. We obtain now amplitudes of 1.32 $\pm$ 0.13 m~s$^{-1}$ for the signal at 10.35 days and 1.30 $\pm$ 0.13 m~s$^{-1}$ for the signal at 21.2 days, with well defined phases. The phase of the sinusoidal signal at 21.20 days remains unchanged. The shorter period signal converges to a period of 10.35 days. The model has a measured lnZ of --764.1 $\pm$ 0.4. This corresponds to a $\Delta$ lnZ of +12.8 with respect to the model using the GP and 1 sinusoid, which corresponds to a 0.0003\% false alarm probability for the more complex model, and a $\Delta$ lnZ of +20.2 against the activity-only model. The GP component remains very similar to the one obtained for the case with 1 sinusoidal, both in RV and FWHM. We measure a rotation period of 127 $\pm$ 15 days, which we adopt as final result. The residuals after the fit in the RV show an RMS of 1.3 m~s$^{-1}$ (0.4 m~s$^{-1}$ for ESPRESSO, 1.6 m~s$^{-1}$ for CARMENES), with no significant signals in their periodogram. Figure~\ref{model_rv_2p} shows the best fit to the data along with the residuals and their periodogram. Table~\ref{parameters_1} shows the priors used and the parameters measured (column `2 Planets (Circ)'). 

We explored the possibility of additional signals being present in the data. Gaussian processes can in many situations suppress signals, in particular at low frequencies. We tested for the presence of a third sinusoidal signal, this time without prior information extracted from the periodograms. We perform two independent tests, one for periods between 22 and 100 days (between the outer sinusoidal and the rotation), and a second one for periods between 100 and 1000 days (longer than the rotation period). In both cases we defined log-uniform priors for the periods and uniform priors between 0 and 1 for the phase of the signal. We do not find any significant signal in the data. Every power excess in the posterior can be explained as artefacts of the sampling. 

We settled for two signals, with periods of 21.20 and 10.35 days. Finally, we tested for the possibility of the signals being eccentric by including two Keplerian models (eq.~\ref{eq_kepler}). We measure a lnZ of --764.5 $\pm$ 0.4. This corresponds to a $\Delta$ lnZ of --0.3 with respect to the model using sinusoidal functions, which favours the circular model. We measure eccentricities of 0.064$^{0.096}_{-0.047}$ for the 21.2-day signal and 0.046$^{0.078}_{-0.035}$ for the 10.35-day signal, consistent with circular in both cases. The periods, amplitudes and times of conjunction of both planets remain consistent with the circular solution. Table~\ref{parameters_1} shows the priors used and the parameters measured (column `2 Planets (Kep)').

\begin{figure*}[ht]
	\includegraphics[width=18cm]{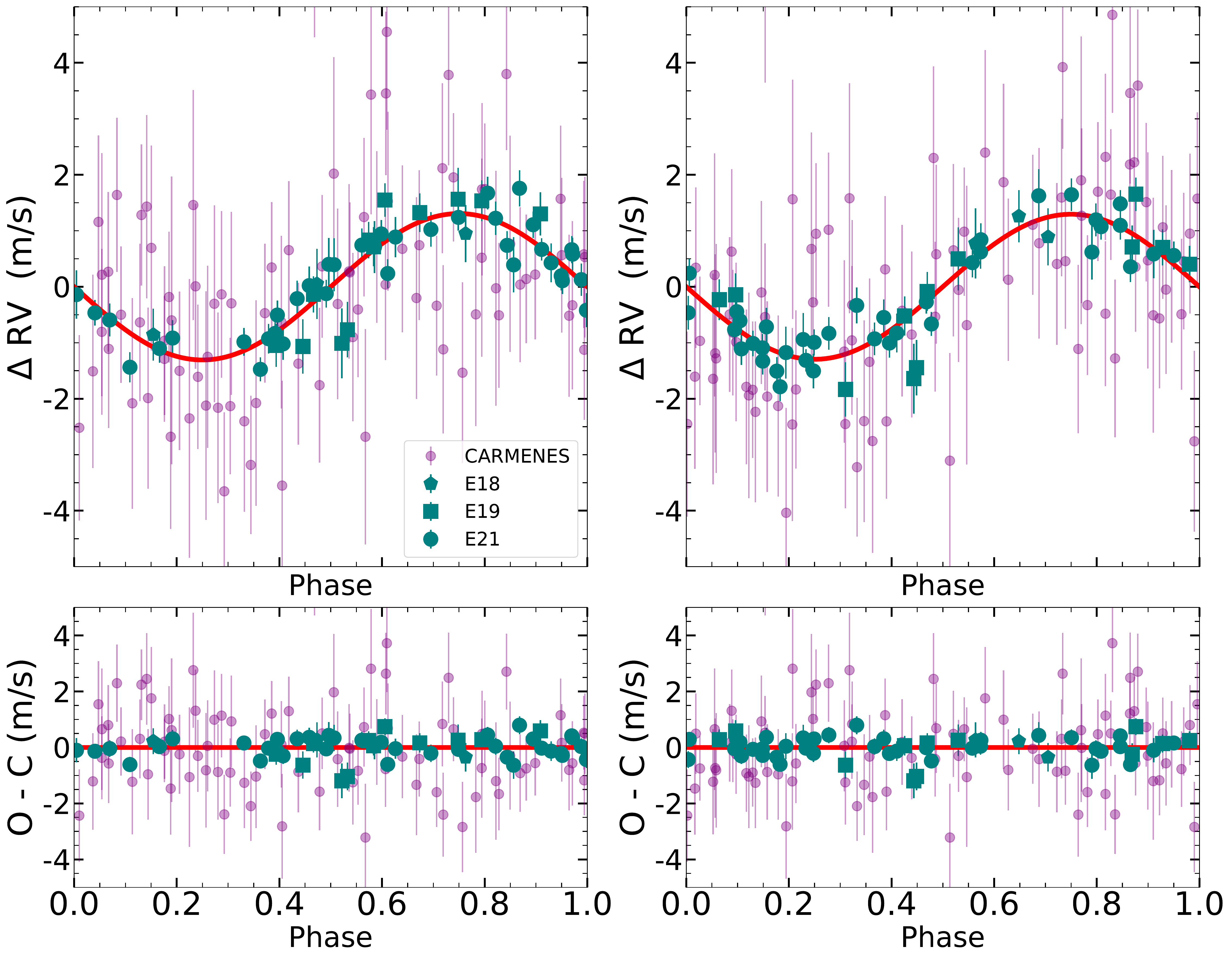}
	\caption{Planetary signals. \textit{Top left panel}: RVs of GJ~1002 b phase folded with a period of 10.35 days, after subtracting the 
        activity model and the signal at 21.2 days. \textit{Top right  panel}: RVs of GJ~1002 b phase folded with a period of 21.2 days, after subtracting the activity model and the signal at 10.335 days. \textit{Bottom panels}: Corresponding phase folded residuals.}
	\label{phase_fold}
\end{figure*}

Given the low eccentricities measured (consistent with zero in both cases) we adopted the model with two circular orbits as our definitive model. We significantly detect the presence of two signals, one with a period of 10.3461$^{ 0.0027}_{-0.0025}$ days, a semi-amplitude of 1.30 $\pm$ 0.14 m~s$^{-1}$ and a $T_{0}$ of 2459566.17 $\pm$ 0.15 d, the other with a period of 21.202$\pm$ 0.013 days, a semi-amplitude of 1.30 $\pm$ 0.14 m~s$^{-1}$ and a $T_{0}$ of 2459560.78 $\pm$ 0.43 d. Figure~\ref{phase_fold} shows the phase folded plots of the two signals, after subtracting the activity model and the other signals, along with the phase folded residuals. We do not identify any structure in the residuals. Figure~\ref{posterior_2pc} shows the posterior distribution of all the parameters sampled in our favoured model. 

\begin{figure}[ht]
	\includegraphics[width=9cm]{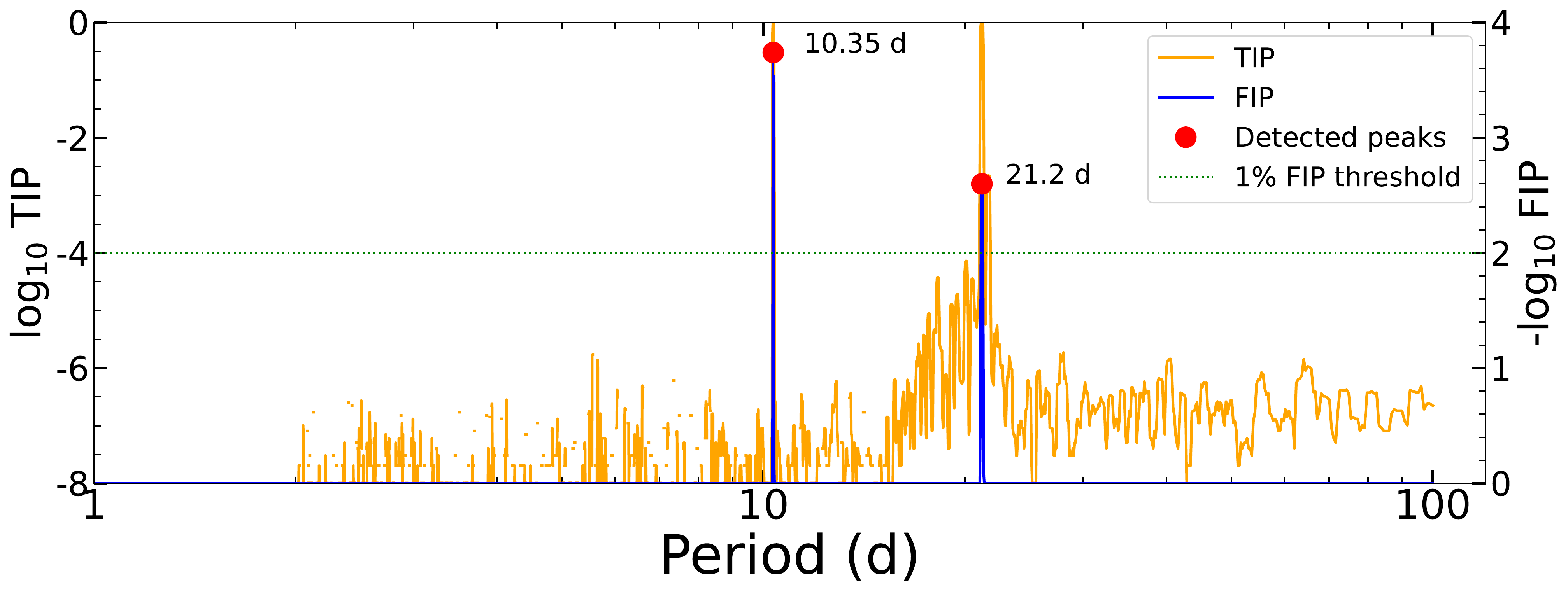}
	\caption{FIP periodogram of the data of GJ~1002. The periods of the peaks are indicated in red points, the $- \rm log_{10} FIP$ and $\rm log_{10} TIP$ are represented as a function of the centre of the period interval considered, in blue and orange respectively. The green dashed line shows the 1\% FIP threshold.}
	\label{fip_tip}
\end{figure}

Additionally, we performed a blind fit in which we used log-uniform priors for the planetary periods, with a range between 2 and 100 days. We recovered consistent results for all parameters, and a $\Delta$ lnZ of +18 with respect to the activity-only model. The larger prior volume required a 4 times larger number of live points for a complete exploration of the parameter space and a much longer computing time.  Taking advantage of the results of the blind fit, and to further test the presence of the planets, we assessed their significance with their False inclusion probability (FIP)  \citep{hara2022a}, which has been shown to be an optimal exoplanet detection criterion \citep{hara2022b}. We consider a grid of frequency intervals, The element $k$ of the grid is defined as $[\omega_k - \Delta \omega/2, \omega_k + \Delta \omega/2]$, where $ \Delta \omega = 2\pi/T_\mathrm{obs}$, $T_\mathrm{obs}$ is the total observation time span, and  $ \omega_k = k\Delta \omega / N_\mathrm{oversampling}$. We take $N_\mathrm{oversampling} = 5$. For each index $k$, we compute the posterior probability to have a planet with an orbital frequency in the interval  $[\omega_k - \Delta \omega/2, \omega_k + \Delta \omega/2]$, or true inclusion probability (TIP), and compute FIP = 1-TIP. In Fig.~\ref{fip_tip}, we represent $-\log_{10} \rm FIP$ as a function of the interval centre. The data show two peaks corresponding to the probability of presence of a planet in the intervals  $[2\pi / 10.341, 2\pi / 10.356]$ and $[2\pi / 21.187, 2\pi / 21.236]$ with probabilities $\textgreater$ 99.99\% and $\textgreater$ 99.7\% respectively.

\subsection{ESPRESSO and CARMENES}

The ESPRESSO and CARMENES data were taken over significantly different periods. Most of the CARMENES data were taken during 2017 and 2018, while most of the ESPRESSO data were taken during 2020 and 2021. If the signals were due to activity, these datasets could show different amplitudes and/or phases. We repeated the analysis by splitting the data in two subsets, one with only ESPRESSO RVs and FWHMs and one with only CARMENES RVs and FWHMs. The lower information content of the independent datasets forced us to consider narrower priors for the periods, of 21.2 $\pm$ 2.1 and 10.4 $\pm$ 1.0. 

\begin{figure*}[ht]
	\includegraphics[width=18cm]{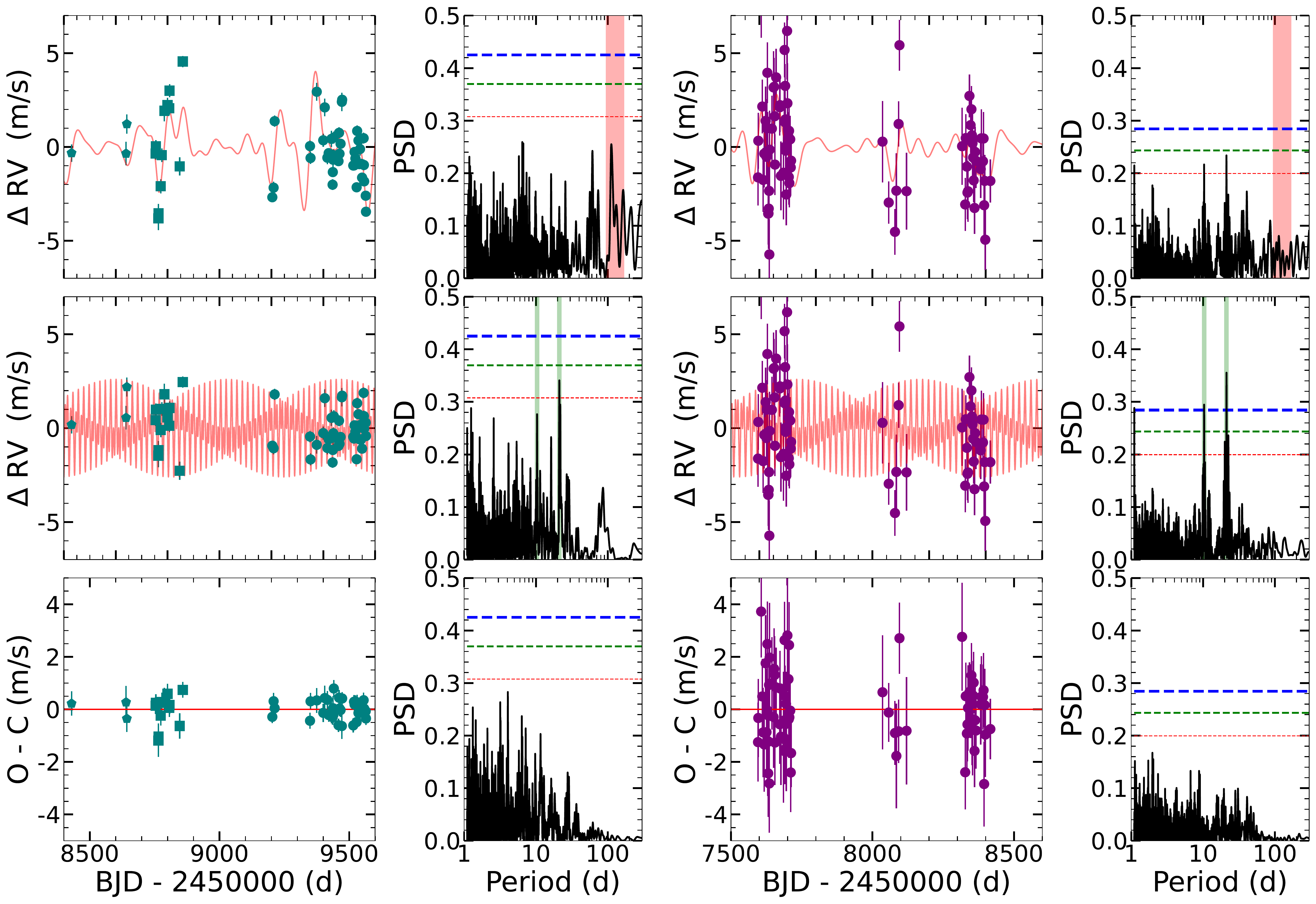}
	\caption{Summary of the periodogram analysis of the independent ESPRESSO and CARMENES datasets. \textit{Top panels}: Raw RVs with the best fit activity model (red line) for ESPRESSO (left) and CARMENES (right), and their respective periodograms. The shaded red region highlights the period interval of the measured stellar rotation. \textit{Middle panels}: RVs, after subtracting the activity model, with the best fit planetary model (red line) for ESPRESSO (left) and CARMENES (right), and their respective periodograms. The shaded green lines highlight the position of the planetary signals identified in the combined analysis. \textit{Bottom panels}: Residual RVs after the best fit (red line) for ESPRESSO (left) and CARMENES (right), and their respective periodograms.}
	\label{esp_car0}
\end{figure*}

\begin{figure}[ht]
	\includegraphics[width=9cm]{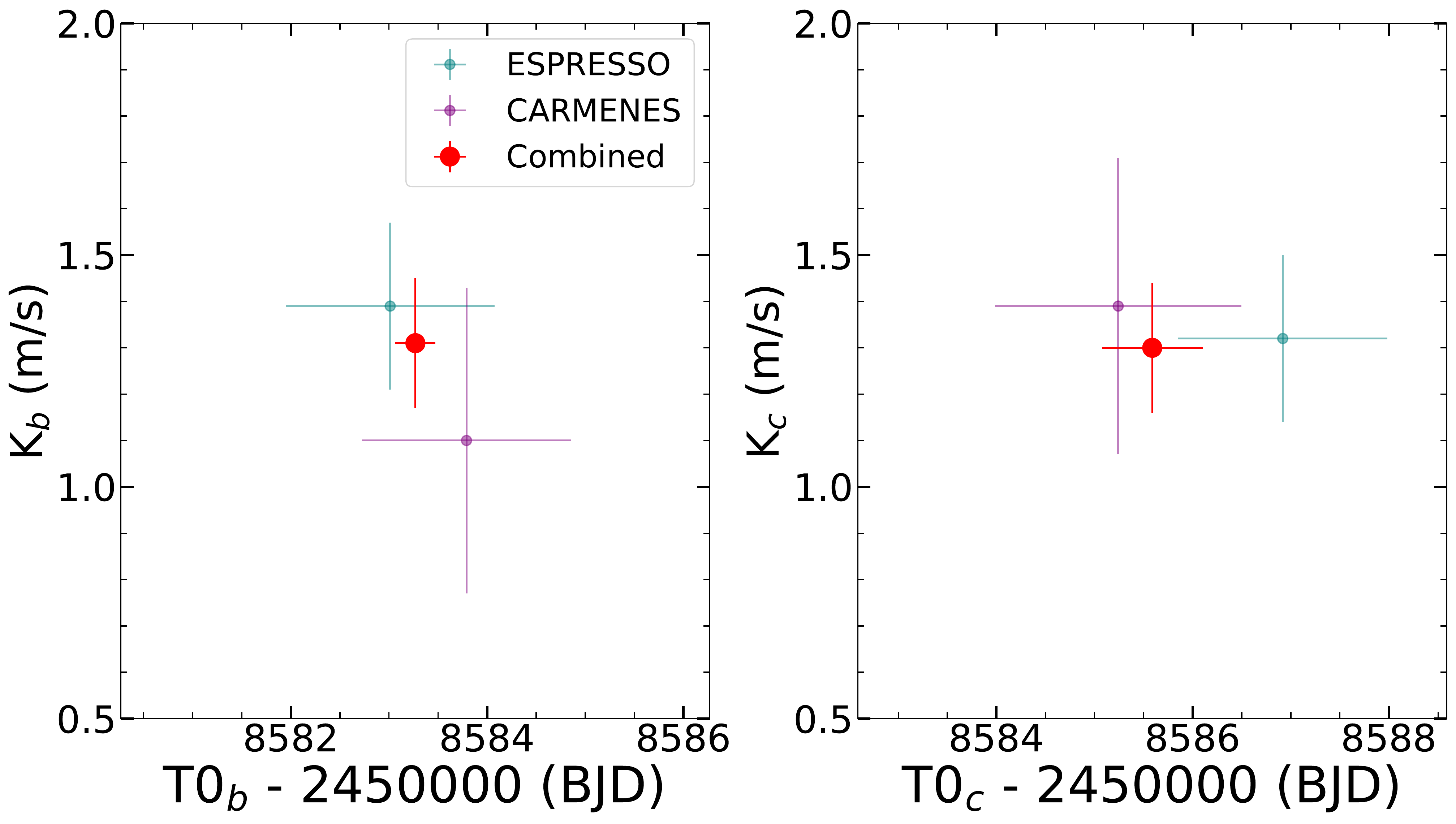}
	\caption{Amplitude versus phase of the two candidate planetary signals for the ESPRESSO, CARMENES and combined dataset.}
	\label{esp_car}
\end{figure}

We obtained very similar results with both instruments as with the combined dataset, although with less significant detections. Figure~\ref{esp_car0} shows a summary of the periodogram analysis performed on the dataset of each instrument. In both cases we obtain strong hints of the presence of the planets. When performing the Bayesian analysis we find that, for both subsets of data, the 2-planet model is significantly favoured over their respective activity-only model. The parameters obtained for the 2-planet model show good consistency between the combined model, the ESPRESSO-only model and the CARMENES-only model. Figure~\ref{esp_car} shows the amplitude versus phase plots of both sinusoidal signals for the ESPRESSO and CARMENES datasets, and for the combined dataset. Additionally we extracted the ESPRESSO RVs using the \texttt{S-BART} algorithm \citep{Silva2022} and repeated the ESPRESSO-only analysis. The results were consistent with those of the \texttt{SERVAL} RVs. 

The parameters obtained in the three datasets are consistent within their error bars. The parameters obtained with the ESPRESSO data are much more precise, dominating the combined fit, but are in all cases within 1$\sigma$ of the parameters derived from the CARMENES data. In spite of the higher quality of the ESPRESSO data, the signals are easier to detect in the periodograms of the CARMENES dataset. The signal caused by the stellar rotation shows a smaller amplitude in the CARMENES data, as expected because of its redder spectral range, making it less prominent with respect to the planetary signals. The global analysis partially circumvents this problem by including a simultaneous fit of the stellar activity component. The observation cadence of the CARMENES campaign is also more appropriate for the detection of short period signals, with a typical spacing between observations (within a campaign) of 2 days, compared to 5 days for the ESPRESSO campaigns. On average, the CARMENES observations sample the shortest planetary signals 5 times per orbit, while the ESPRESSO data sample it 2 times per orbit. The effect of the lower cadence can be seen in Figure~\ref{phase_fold}, where there ESPRESSO data still show some gaps in their phase coverage. 

\subsection{Alternative activity models}

We significantly detected the presence of two planetary signals. Since no activity signals are detected at their periods, and considering the low activity of the star, it seems unlikely that the signals are due to stellar activity. However, it is possible that an incorrect activity model could leave some of the harmonic components of the activity signal unmodelled, or could create artefacts at periods shorter than the stellar rotation. To double-check whether or not the detected signals could be caused by the activity modelling, we repeated the process with several other GP kernels. We tested the Mat\'ern 3/2 exponential periodic (MEP) and exponential-sine periodic (ESP) kernels, provided by the \texttt{S+LEAF} package. These two kernels are approximations of the classical quasi-periodic GP Kernel \citep{Haywood2014}. This approach, as our main one, works on a multi-series GP building an activity model as a linear combination of the GP and its derivative. We also analysed the data using a GP modelling without the derivative. We used the double SHO and the pseudo quasi-periodic kernel with a component at half the rotation period (PQP2) \citep{Masca2021}, built using the \texttt{celerite2} package, and the quasi-periodic Kernel, built using the \texttt{GEORGE} package \citep{Foreman-Mackey2014}. In all cases we analysed the data with and without simultaneous modelling of the FWHM. Lastly, we re-analysed the data using a simpler model for the stellar activity, based on two sinusouds at the rotation period and its second harmonic. We did not obtain significantly different results with any of the algorithms tested. All the models provided significant detections and consistent parameters for the two candidate planetary signals, although with different noise levels and uncertainties. After all our tests we find it unlikely that the detected signals are unmodelled harmonics of the rotation or artefacts of the modelling process. 

\subsection{Stability of the signals} \label{testing}

\begin{figure}[ht]
	\includegraphics[width=9cm]{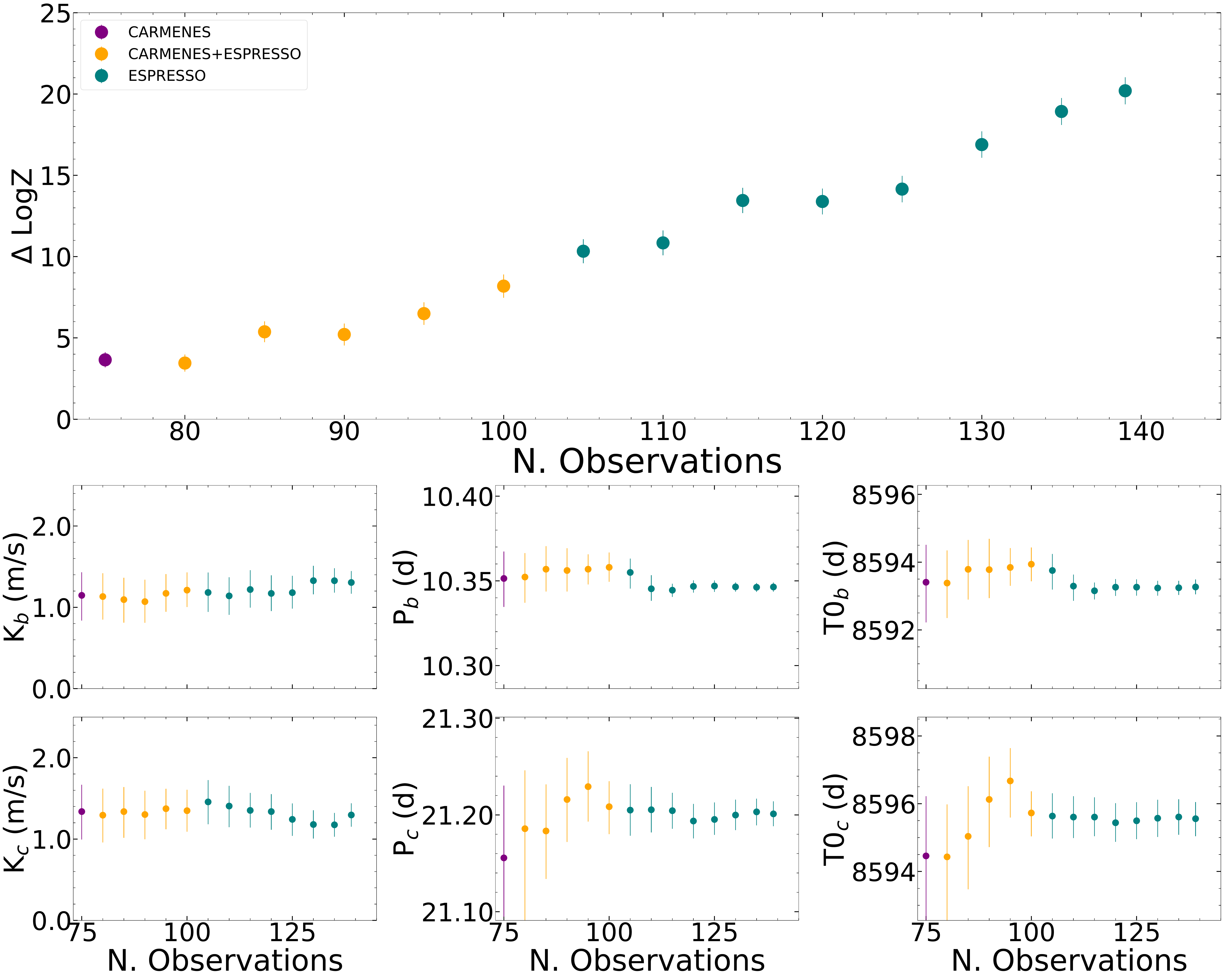}
	\caption{Signal stability test. \textit{Top panel}: Evolution of the difference in Bayesian evidence ($\Delta$ lnZ) of the 2-planet model against an activity-only model. \textit{Bottom panels}: Evolution of the parameters of the two sinusoids. The colours of the points show which data are added at each step.}
	\label{signals_evol}
\end{figure}

Keplerian signals are expected to be stable in RV datasets, while activity signals are not, due to their lack of long-term coherency, \citep{Mortier2017, Masca2018a}. We study the behaviour of our two detected signals as a function of the number of measurements and follow the evolution of their significance and measured parameters. If the signals correspond to planets, we expect a steady increase in the significance of the detection ($\Delta$ lnZ of the 2-planet model vs an activity-only model), and their parameters to converge towards the final value and remain stable once we reach sufficiently high signal to noise ratio. While this method is based on the intrinsic differences between planetary signals and stellar activity, it has to be noted that the evolution of all the measured parameters is also affected by extrinsic phenomena such as observation cadence and by the quality of the data. Stellar activity can also have a non-negligible effect on the parameters of planetary signals, by increasing the noise levels or affecting the fit if the activity and planetary amplitudes are similar. Adding new instruments can also reduce the significance when needing to fit for the offset and jitter values with very few points. Figure~\ref{signals_evol}, top panel, shows the evolution of the Bayesian evidence of the two-planet model against an activity-only model, as we include more data in order of time. Once the signals are detected (at $\sim$ 80 measurements), the evolution of the significance of the model is very steady, suggesting that the two sinusoidal signals do not have their origin in stellar activity. On a similar note, their parameters (semi-amplitude, period and phase; Fig~\ref{signals_evol}, bottom panels) evolve smoothly towards their final value, showing very little variation after $\sim$ 100 measurements. There are some anti-correlated variations between both amplitudes (within the 1$\sigma$ range) that are most likely caused by the coupling of their near-resonance periods and the observational cadence. It is important to note that the only campaign with sufficiently high cadence to separate both signals in a straightforward way was the latest 2021 ESPRESSO campaign. The steady evolution of the significance and the stability of the recovered parameters support a planetary origin for the two signals. 

Our tests show the two signals to be stable over long periods of time and to be independent of the analysis procedure. Our previous analysis of stellar activity provided a good characterisation of the stellar signals present in the data, which was perfectly consistent with typical activity-rotation relationships, and showed no short period activity signals close to the planetary signals. With this information, we conclude that the signals at 10.35 and 21.2 days are indeed most likely induced by low-mass planets orbiting the star. 

\section{Discussion}

\subsection{Planetary system}

We detected the significant presence of two planetary signals. GJ~1002 b has an orbital period of 10.3461$^{ 0.0027}_{-0.0025}$ and its RV signal has a semi amplitude of 1.32 $\pm$ 0.13 m~s$^{-1}$. These values correspond to a planet with a minimum mass of 1.09 $\pm$ 0.13 M$_{\oplus}$ orbiting at a distance of 0.0457 $\pm$ 0.0013 au of its parent star. GJ~1002 c has an orbital period of 21.202$\pm$ 0.013 days and its RV signal has a semi amplitude of 1.30 $\pm$ 0.13 m~s$^{-1}$. These values correspond to a planet with a minimum mass of 1.36 $\pm$ 0.16 M$_{\oplus}$ orbiting at a distance of 0.0738 $\pm$ 0.0021 au of its parent star. 

Using the ephemeris derived from the RV analysis, we performed an informed transit search in the \textit{TESS} light curve. Our results predict two potential transit events of GJ~1002 b and no transit for GJ~1002 c. We did not detect any transit event in the TESS data. Assuming a planet radius of $\sim$ 1.1 R$_{\oplus}$ for GJ~1002 b, the expected transit depth would be of $\sim$ 5 ppt. The dispersion of the light curve excludes a transit event. We cannot derive strong limits for the inclination of the system with respect to the line of sight, as only planets orbiting with an inclination $i$ > 89.2$^\circ$ are expected to transit at the orbital distance of GJ~1002 b.

Following \citet{Kopparapu2017} we estimate that both planets lie within the HZ of the star (see Sect.~\ref{stellar_params}). GJ~1002 b receives an incident flux S $\sim$ 0.67 $F_{\oplus}$ and GJ~1002 c receives S $\sim$ 0.26 $F_{\oplus}$. With these values of incident flux, both planets are comfortably within the conservative HZ for their star \citep{Kopparapu2017}. We also estimate their equilibrium temperature, assuming an albedo of 0.3. We calculate equilibrium temperatures of 230.9 K for GJ~1002 b and 181.7 K for GJ~1002 c, slightly cooler than Earth and Mars, respectively.

\begin{figure*}[ht]
	\includegraphics[width=18cm]{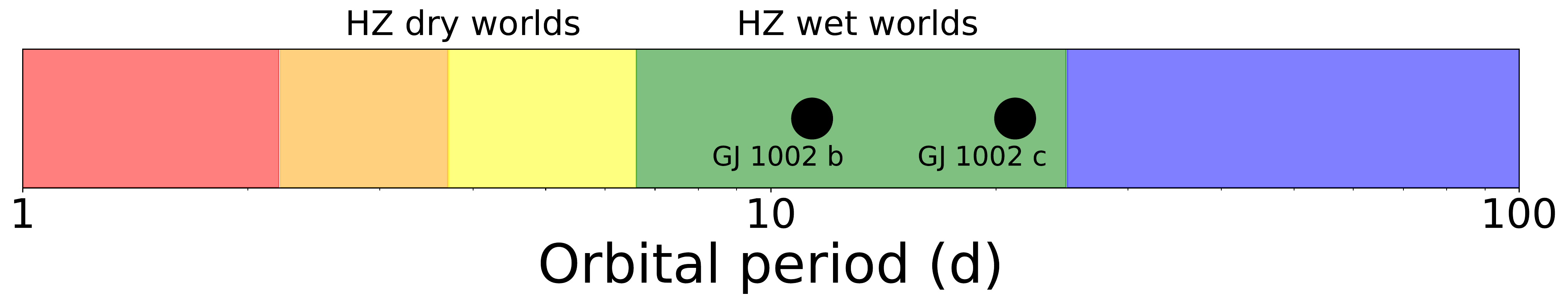}
	\caption{Position of the planets detected orbiting GJ~1002, with respect of the limits of the HZ. The orange and yellow region show the HZ for dry worlds \citep{Zsom2013}, while the green region shows the HZ for wet worlds \citep{Kopparapu2017}. The red and blue regions show the regions in which it would be too hot or two cold.}
	\label{habit}
\end{figure*}

Table~\ref{param_planets} summarises the parameters of both planets and Fig.~\ref{posterior_2pc} shows the posterior distribution obtained for the planet parameters in the model with two circular orbits. With GJ~1002 located at just 4.84 pc from the Sun, it is one of the closest known multi-planetary systems hosting temperate Earth-mass exoplanets. GJ~1002 joins Proxima \citep{AngladaEscude2016}, Ross 128 \citep{Bonfils2018}, GJ 1061 \citep{Dreizler2020} and Teegarden's star \citep{Zechmeister2019} as systems with planets that could potentially host habitable environments within 5 pc. Figure~\ref{habit} shows the configuration of the system, with respect to the limits of the HZ. 

It should be noted that it is not fully clear whether low-mass M-dwarfs can host habitable planets. They produce frequent and intense flares during their pre-main sequence evolution and well into their journey through the main sequence (e.g \citealt{Davenport2016}), and have very low quiescent stellar fluxes. It has been suggested that the intense flare activity could strip the planets from their atmosphere, preventing the development of life (e.g \citealt{Howard2018}). It is also possible that planets in their habitable zone do  not receive enough ultra-violet radiation to start life via photochemistry. However, the energy received via strong flaring during the early age of the star might be sufficient to build up the prebiotic inventory. It might also not be too much of a problem for the continuation of life if the energy of those flares decreases significantly with age \citep{Rimmer2018}. Given its mass, it is expected that GJ~1002 showed an intense flaring activity during its youth and rotational breaking phase \citep{Mondrik2019}, but that activity seems to have ceased long ago, as \textit{TESS} observed no flares. GJ~1002 also shows very weak X-ray emission, compared to other low-mass M-dwarfs \citep{SchmittLiefke2004}. If its presumed past activity was strong enough to trigger photochemistry on the surface of its planets, its current gentle activity could be low enough to not threaten the stability of RNA strands.

\begin{table} 
\begin{center}
\caption{Parameters of the two planets detected orbiting GJ~1002 using under the assumption of circular orbits \label{param_planets}}
\begin{tabular}[centre]{l c c}
\hline
Parameter  & GJ~1002 b & GJ~1002 c \\ \hline
$T_{0}$ -- 2450000 [d] & 8583.27 $\pm$ 0.22 &  8585.54 $\pm$ 0.50  \\
$P_{\rm orb}$ [d] & 10.3465 $\pm$ 0.027 & 21.202 $\pm$ 0.013 \\
$K_p$ [m~s$^{-1}$] &  1.31 $\pm$ 0.14 & 1.30 $\pm$ 0.14\\
$m_p \sin i $  [M$\oplus$] & 1.08 $\pm$ 0.13 & 1.36 $\pm$ 0.17\\
$a$ [au] &  0.0457 $\pm$ 0.0013 & 0.0738 $\pm$ 0.0021\\
Incident flux [$F_{\oplus}$] & 0.670 $\pm$ 0.039 & 0.257 $\pm$ 0.015 \\
Eq. Temp.$_{A = 0.3}$ [K] & 230.9 $\pm$ 6.7 & 181.7 $\pm$ 5.2 \\
\hline
\end{tabular}
\end{center}
\end{table}

\begin{figure*}[ht]
	\includegraphics[width=18cm]{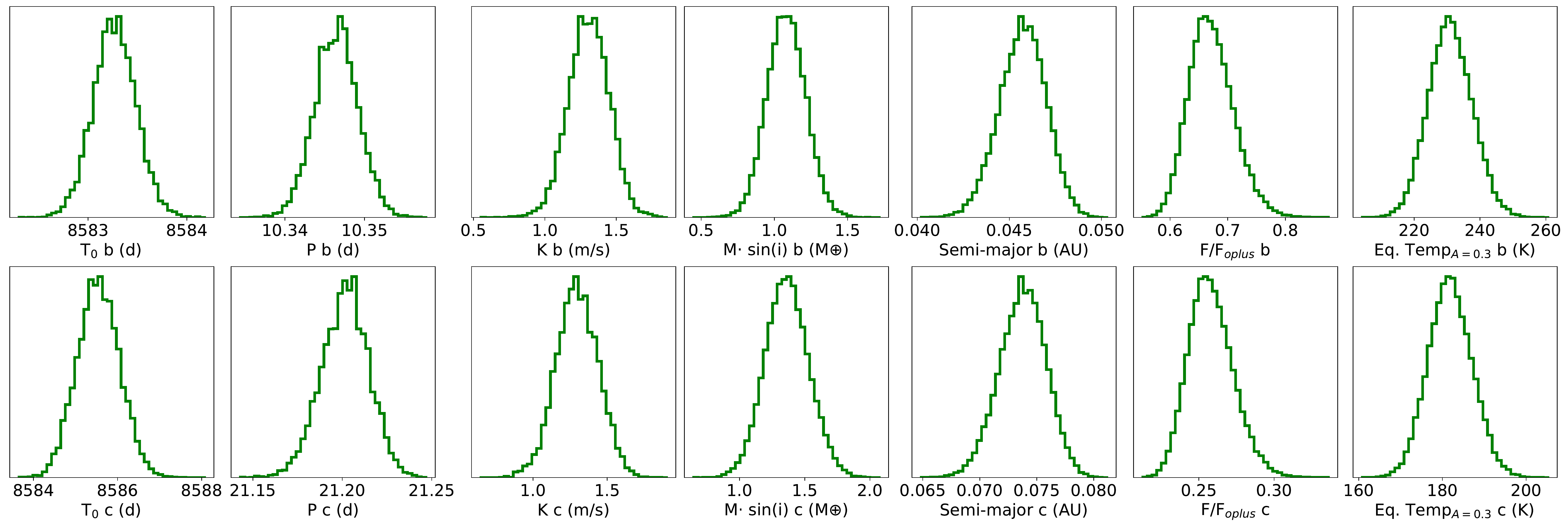}
	\caption{Posterior distribution (green line) of the planetary parameters sampled in the favoured model in section~\ref{analysis_planets}, along with their derived parameters. The red line shows the distribution of priors.}
	\label{posterior_pl}
\end{figure*}

\subsection{Prospects for atmospheric characterisation}

Atmospheric characterisation of exoplanets is typically performed via transmission spectroscopy, which requires the planets to transit. However, that is not the only possibility to study the atmospheres of exoplanets. In recent years it has been proposed that coupling high-resolution spectrographs with high angular resolution imaging could open an alternative path to studying the atmospheres of the planets in the solar neighbourhood \citep{Lovis2017}. In the coming years, RISTRETTO, a proposed visitor instrument for the VLT, will enable the possibility of performing high dispersion coronagraphy (HDC) spectroscopy of exoplanets, which is expected to allow the study of exoplanet atmospheres at contrasts of 10$^{-7}$ at angular distances of 2 times wavelength over telescope diameter ($2~\lambda/D$) \citep{Blind2022}. The aim is to detect the oxygen band at 760 nm, where the VLT angular distance $2~\lambda/D$ is 37 mas. RISTRETTO will serve as a proof of concept for ANDES \citep{Marconi2021}, the high-resolution spectograph for the ELT. ANDES will be a R$\sim$ 100~000 stabilised \'Echelle spectrograph with a wavelength coverage of at least 400 -- 1800 nm. It will include an integral field unit fed by a single-conjugate adaptive optics (SCAO) module in the NIR part of the spectrograph (950 -- 1800 nm) to correct for the blurring effect of turbulence in the atmosphere\footnote{https://elt.eso.org/instrument/ANDES/}. It is expected to perform HDC spectroscopy of exoplanets, as RISTRETTO, with the goal of detecting biomarkers. An examples of those biomarkers is the oxygen band at 1300 nm. The angular separation of the orbits of GJ~1002 b and c are $\sim$ 9.7 mas and $\sim$ 15.7 mas, respectively. Following \citet{Lovis2017} we estimate the planet-star flux ratio to be 10$^{-7}$ and 7 $\cdot$ 10$^{-8}$, respectively. The distance $2~\lambda/D$ for the ELT at 1300 nm is $\sim$ 13.7 mas (GJ~1002 c within reach). Using ANDES, it should be possible to probe the presence an atmosphere on the outer planet and test the presence of oxygen in its atmospheres. 

Another possible path to study their atmospheres is the measurement of their thermal emission. LIFE \citep{Quanz2022} is a proposed mid-infrared nulling interferometry space mission designed to measure the thermal emission of exoplanets. It is expected to consist in four 2-metre apertures working in the 4--15.5~$\mu$ wavelength range. With incident fluxes of 0.67 and 0.257 $F_{\oplus}$, and expected radii of 1--1.5~$R_{\oplus}$, both planets of GJ~1002 fall within the range the mission expects to study \citep{Quanz2022}.

\subsection{Detection limits for additional companions}

We detected the presence of two Earth-mass planets orbiting GJ~1002. However, that does not mean there are no other planets orbiting GJ~1002. The quality of the data, observation strategy, activity of the star and activity modelling all affect our detection sensitivity. 

We estimate our detection limits for additional companions by performing a simple injection-recovery test. We construct activity-only RVs by subtracting the two planetary signals from figure~\ref{phase_fold}. We then inject 100,000 random sinusoidal signals with periods between 1 and 1000 days, amplitudes between 1 cm~s$^{-1}$ and 10 m~s$^{-1}$, and random phases. We reject those combinations that would create an RMS of the data higher than the real RMS of our data. Then we subtract the stellar activity using the same GP hyper-parameters as measured for our 2-planet model, but recomputing the activity model for each specific dataset. Then we generate the periodogram of the residuals and check the false alarm probability of the highest peak at the injected period, in the same way we did for our original activity-only model (Fig.~\ref{model_rv_0p}). 

\begin{figure}[ht]
	\includegraphics[width=9cm]{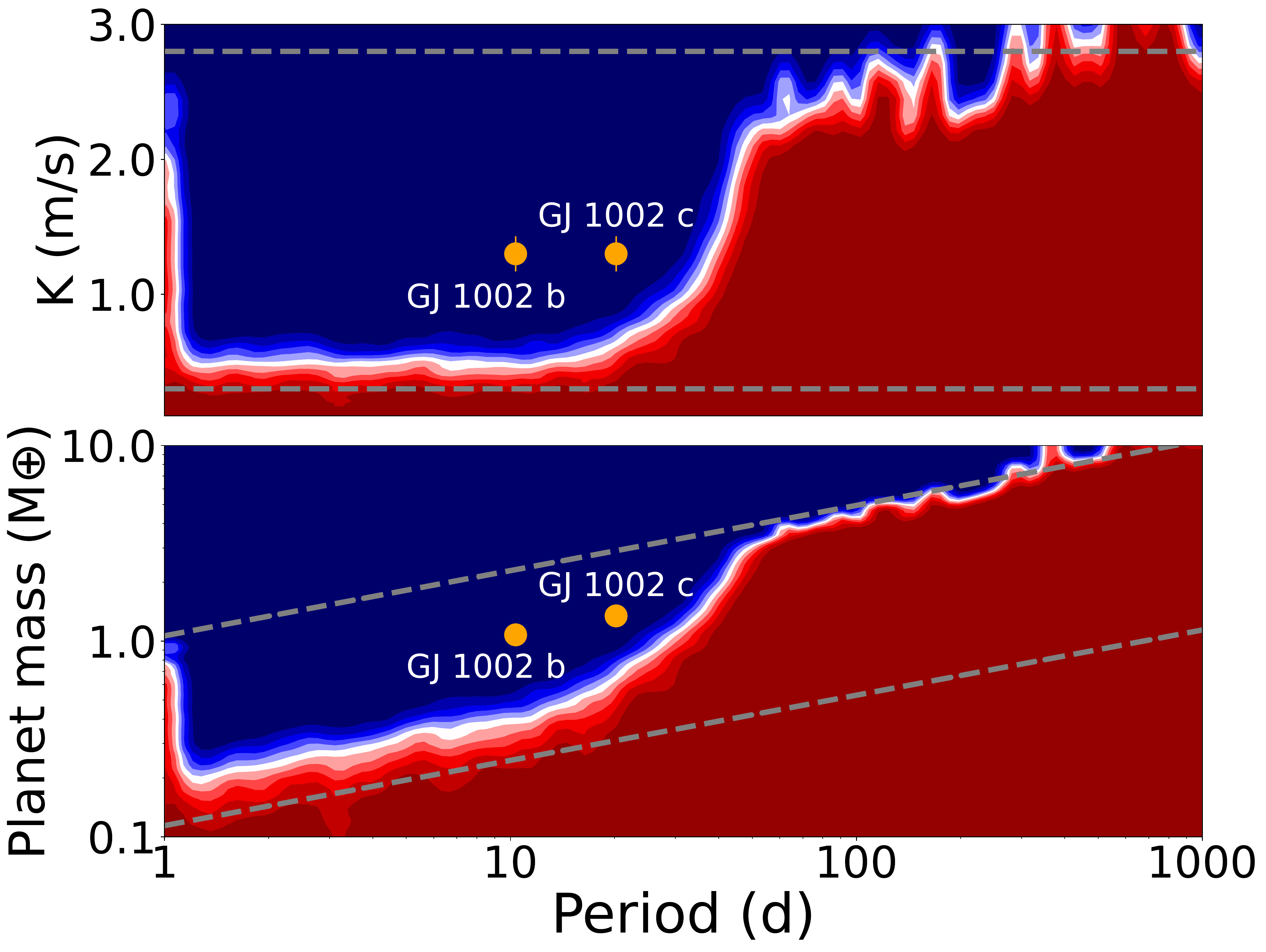}
	\caption{Detection limits. \textit{Top panel}: Amplitude versus period diagram for our detection limits based on injection and recovery tests. The white region shows the signals that we would detect in a periodogram with 1$\%$ FAP. The blue region shows signals we would detect with a FAP lower than 1$\%$ FAP. With the current dataset and activity model, we would miss the red region signals. The top grey dashed line shows the maximum amplitudes that would still be compatible with the data. The bottom grey dashed line shows the minimum amplitude that we expect to detect with ESPRESSO data and the current exposure times. The orange circles show the position of the detected planets. \textit{Bottom panel}: Same as top panel but in mass versus period. }
	\label{det_lims}
\end{figure}

Figure~\ref{det_lims} show the result of this exercise. With our current dataset and activity model, we are sensitive to signals of $\sim$ 0.5 m~s$^{-1}$ for periods from 1.2 days up to $\sim$ 15 days. Our sensitivity drops to $\sim$ 1 m~s$^{-1}$ at $\sim$ 35 days and then quickly plummets to $\sim$ 2 m~s$^{-1}$ at periods longer than $\sim$ 40 days. These values correspond to masses of $\sim$0.5 M$_{\oplus}$ at 15 days, $\sim$1 M$_{\oplus}$ at 35 days and > 3 M$_{\oplus}$ at periods longer than 40 days. This is a typical effect of GP-only models, which try to account for all variations at low frequencies up to the characteristic value specified by the kernel. To extract possible low-amplitude signals at longer periods we would need either a very significant amount of new data or a different strategy to mitigate stellar activity. This also shows that finding the signals of GJ~1002 b and c in the periodogram of the residuals is within expectations. 

With these results, we do not expect to have missed any Earth-like planets in the HZ. The sensitivity that we achieve in our exercise is, for short periods, very close to the RV precision. To explore the region of lower mass planets at short orbits we anticipate a change in strategy would be necessary, very likely increasing the exposure time of the observations to reduce the photon noise uncertainty of the RV measurements, for what might be marginal gains. Exploring the range between 40-130 days would be a more interesting, and maybe less time-consuming, prospect. There is still plenty of room around GJ~1002 for additional Earth-like planets (or super-Earths). If they exist, these planets would also be prime targets for characterisation with future instruments, as GJ~1002 c.

The \textit{Gaia} DR3 archive reports astrometric excess noise of 0.36 mas and a Renormalised Unit Weight Error (RUWE) of 1.33. The RUWE is expected to be around 1.0 for sources where the single-star model provides a good fit to the astrometric observations\footnote{https://gea.esac.esa.int/archive/documentation/}. The RUWE value, just below the threshold of 1.4 used by \citet{Arenou2022} for processing sources as non-single stars, could be due to an additional companion, unseen in the RV data. 

\begin{figure}[ht]
	\includegraphics[width=9cm]{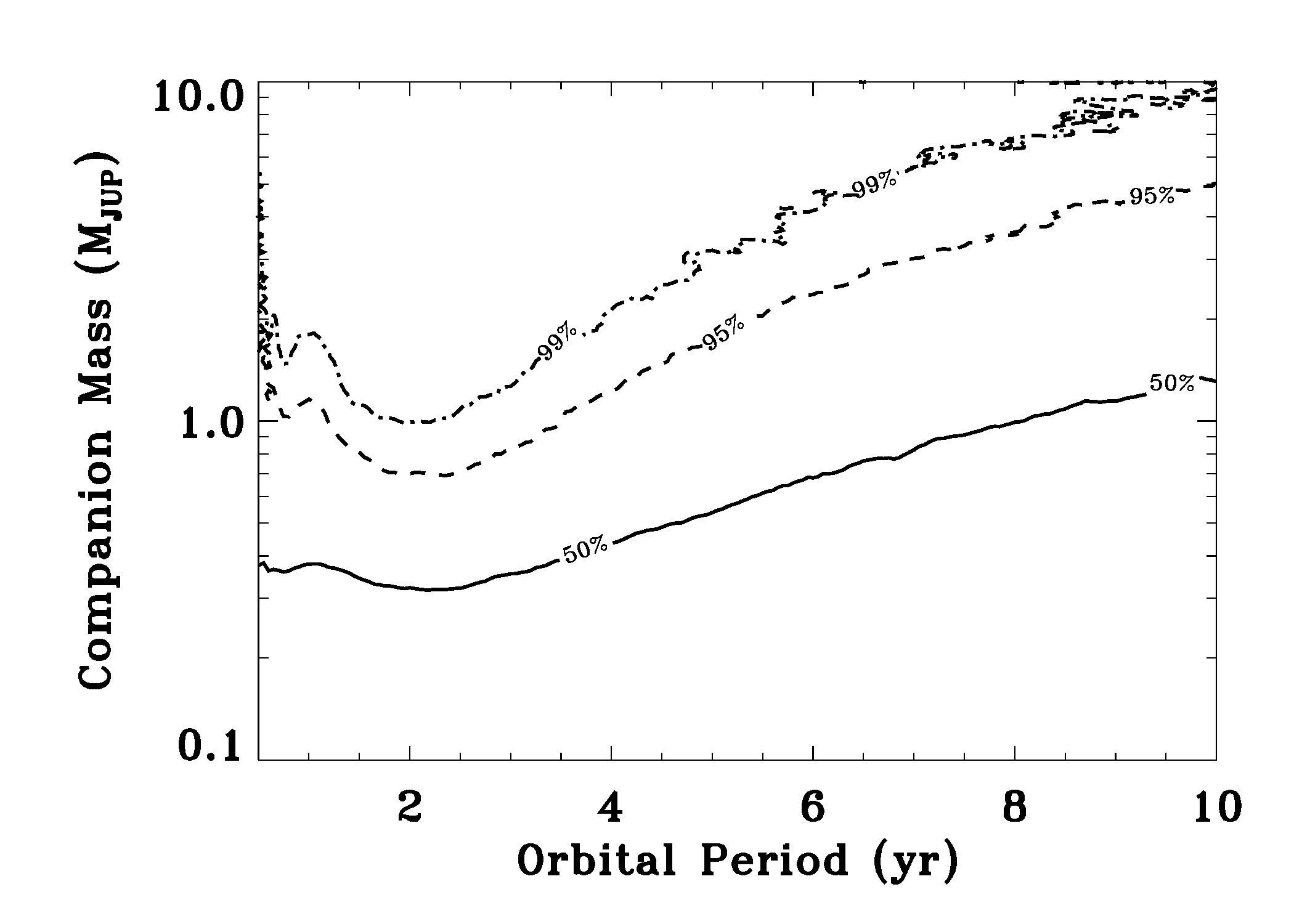}
	\caption{Gaia DR3 sensitivity curves for the star. The curves correspond to a probability of 50\%, 95\% and 99\% of a companion of given mass and period to produce a RUWE larger than the observed one.}
	\label{det_lims_gaia}
\end{figure}

We calculated the sensitivity limits of the \textit{Gaia} DR3 astrometry for GJ~1002. Planetary companions in the range 1--10 M$_{\rm JUP}$ would produce a RUWE in excess of the measured value for periods in the range 0.5--10 yr (Figure~\ref{det_lims_gaia}). For edge-on orbits, and periods up to 8 years, this would correspond to RV signals significantly larger than what we expect to have missed (see. Fig~\ref{det_lims}), possibly pointing to a quasi-face-on orbital configuration for the companion, or an orbital period much larger than our baseline of observations.  

To characterise the full architecture of the planetary system of GJ~1002 it will be necessary to perform additional RV campaigns, with instruments such as ESPRESSO, Maroon-X \citep{Seifahrt2018}. These campaigns need to have their observing strategy tailored to better characterising the activity signal. Another option is to attempt them with NIR spectrographs such as NIRPS \citep{Bouchy2017, Wildi2017}, taking advantage of the expected chromatic behaviour of activity signals. Future astrometric measurements are necessary to establish the origin of the detected astrometric excess noise.

\subsection{Activity of GJ  1002}

GJ~1002 is a low-activity M5.5V dwarf. We measure a rotation period of 126 $\pm$ 15 days, which is consistent with the typical rotation-activity relationships for M-dwarfs \citep{Masca2016, Masca2018}. The amplitude of the activity signal, as measured using a double-sinusoidal model (from Sect.~\ref{testing}) is $\sim$ 0.7 m~s$^{-1}$, which is also consistent with the amplitude-activity relationship of \citet{Masca2018}. We did not find evidence for the presence of a long-term cycle in our data. However, our dataset is not optimal for the detection of long-term cycles. Overall, GJ~1002 is a fairly typical low-activity fully convective M-dwarf. Its behaviour over a few years fits within the expectations based on its typical level of chromospheric activity and is not very different from other old M-dwarfs such as Barnard's star \citep{Toledo-Padron2019}.

\subsection{Temperature of the optical elements and instrumental telemetry}

Most spectrographs designed to obtain precise RVs are thermally and pressure stabilised. This stabilisation ensures that there are no large variations in the internal conditions of the spectrograph, which would otherwise create variations in the RV measurements. On top of that, most modern RV spectrographs use some form of simultaneous calibration which helps correcting the remaining RV variations due to remaining temperature changes. However, we have seen that even very small temperature changes of the optical elements can affect the instrumental profile and induce changes in some of the metrics typically used as activity indicators. 

We modelled the temperature effect by relating the FWHM and \'Echelle temperatures using a third order polynomial. We recovered parameters different from zero at a 1-4 $\sigma$ level, depending on the parameter and the instrument (See table~\ref{parameters_1} and figure~\ref{posterior_2pc}). We conclude that there is indeed a low-amplitude temperature effect that can easily pass unnoticed. These variations can then introduce parasitic signals in the time series that can be easily mistaken for stellar effects. 

We did not find similar variations in the RV time series of GJ~1002. We attempted to link the RV and \'Echelle temperatures in a similar way, but all parameters were consistent with zero for both instruments. However, it is not guaranteed that they will not be present in higher signal-to-noise datasets (brighter stars) or datasets showing a lower RV RMS. 

Most observatories and instruments save enough telemetry data to track these changes in the conditions of the instruments, but these data are rarely analysed (or at the very least this analysis is rarely reported). As we progress towards the detection of very low-mass planets, such as sub-Earths orbiting M-dwarfs, or the detection of Earth-mass planets orbiting solar-type stars, it is important to remember that ultra-stabilised spectrographs are not perfect spectrographs. To detect these planets, it is important to understand better the behaviour of the instrumentation. When the exoplanet community started to detect planets at 1 m~s$^{-1}$ amplitudes, some form of stellar activity analysis became a necessary step. If we aim to detect planets at 10 cm~s$^{-1}$, some form of instrumental telemetry analysis will very likely become unavoidable. 

\section{Conclusions}

We studied the nearby M-dwarf GJ~1002 using RVs and activity indicators from ESPRESSO and CARMENES. Using a joint model that combined information from the FWHM of the CCF and RVs into a multi-series Gaussian process, we detected the presence of two planetary signals. GJ~1002 b has an orbital period of 10.3465$\pm$0.0027 days and a minimum mass of 1.08 $\pm$ 0.13 M$_{\oplus}$. GJ~1002 c has an orbital period of 21.202$\pm$ 0.013 days and a minimum mass of 1.36 $\pm$ 0.17 M$_{\oplus}$. With both planets orbiting within the HZ of their star, GJ~1002 becomes one of the nearby systems with planets that could potentially host habitable environments.

We find a rotational modulation of 126 $\pm$ 15 days, consistent with the expectations from rotation-activity relationships, and no clear evidence for a long period modulation present in the FWHM or the RV time series. As with the case of other M-dwarfs, the FWHM seems to be a good activity indicator for tracking the rotational modulation of the star. We found that the multi-series GP approach provides good results at modelling the data, although the results were not easily distinguishable from a more typical GP modelling.

We rule out the presence of additional companions with masses larger than 0.5 M$_{\oplus}$ at periods shorter than $\sim$ 20 days, and of companions with masses larger than 3 M$_{\oplus}$ at periods up to $\sim$ 50 days. It is still possible that there are additional Earth-mass planets in the outer half of the HZ. The \textit{Gaia} DR3 data show an excess astrometric noise, that could point to a massive companion at large orbital separation.

GJ~1002 c is a good candidate for atmospheric characterisation via High Dispersion Coronagraphy spectroscopy with the future ANDES spectrograph for the ELT. The atmospheres of that planet should be observable around the oxygen absorption at 1300 nm.

We found that long period variations in the temperature of the optical elements of ESPRESSO and CARMENES affect the measurements of the FWHM of the CCF, most likely by changing the instrumental profile. We do not find evidence that these temperature variations affect the RV measurements in our dataset. However, it is not guaranteed that this would be the case for all RV datasets.

\section*{Acknowledgements}
We want to thank the anonymous referee for their insightful comments, which helped improve the manuscript. We want to thank Ravi Kopparapu and Emily Gilbert for reaching out and highlighting an error in the limits of the habitable zone. ASM acknowledges financial support from the Spanish Ministry of Science and Innovation (MICINN) under 2018 Juan de la Cierva program IJC2018-035229-I. ASM, VMP, JIGH, CAP and RR acknowledge financial support from the Spanish MICINN project PID2020-117493GB-I00. ASM, VMP, JIGH and RR acknowledge financial support from the Government of the Canary Islands project ProID2020010129. J. P. F. is supported in the form of a work contract funded by national funds through FCT with reference DL 57/2016/CP1364/CT0005. CJM acknowledges FCT and POCH/FSE (EC) support through Investigador FCT Contract 2021.01214.CEECIND/CP1658/CT0001. This work was supported by FCT - Fundação para a Ciência e a Tecnologia through national funds and by FEDER through COMPETE2020 - Programa Operacional Competitividade e Internacionalização by these grants: UIDB/04434/2020; UIDP/04434/2020. A.M.S acknowledges support from the Fundação para a Ciência e a Tecnologia (FCT) through the Fellowship 2020.05387.BD. and POCH/FSE (EC). This  work  was  supported  by  FCT  - Fundação para   a Ciência   e   a   Tecnologia   through national funds and   by   FEDER through  COMPETE2020  - Programa Operacional Competitividade  e Inter-nacionalização by  these grants: UIDB/04434/2020; UIDP/04434/2020; PTDC/FIS-AST/32113/2017  \& POCI-01-0145-FEDER-032113; PTDC/FIS-AST/28953/2017 \& POCI-01-0145-FEDER-028953. The INAF authors acknowledge financial support of the Italian Ministry of Education, University, and Research
with PRIN 201278X4FL and the "Progetti Premiali" funding scheme. JBD acknowledges financial support from the Swiss National Science Foundation (SNSF). This work has, in part, been carried out within the framework of the National Centre for Competence in Research PlanetS supported by SNSF. CARMENES is an instrument for the Centro Astronómico Hispano en Andaluc\'ia (CAHA) at Calar Alto (Almería, Spain), operated jointly by the Junta de Andaluc\'ia and the Instituto de Astrofísica de Andaluc\'ia (CSIC). CARMENES was funded by the Max-Planck-Gesellschaft (MPG), the Consejo Superior de Investigaciones Científicas (CSIC), the Ministerio de Economía y Competitividad (MINECO) and the European Regional Development Fund (ERDF) through projects FICTS-2011-02, ICTS-2017-07-CAHA-4, and CAHA16-CE-3978, and the members of the CARMENES Consortium (Max-Planck-Institut für Astronomie, Instituto de Astrofísica de Andaluc\'ia, Landessternwarte Königstuhl, Institut de Ciències de l’Espai, Institut für Astrophysik Göttingen, Universidad Complutense de Madrid, Thüringer Landessternwarte Tautenburg, Instituto de Astrofísica de Canarias, Hamburger Sternwarte, Centro de Astrobiología and Centro Astronómico Hispano en Andaluc\'ia), with additional contributions by the MINECO, the Deutsche Forschungsgemeinschaft through the Major Research Instrumentation Programme and Research Unit FOR2544 “Blue Planets around Red Stars”, the Klaus Tschira Stiftung, the states of Baden-Württemberg and Niedersachsen, and by the Junta de Andaluc\'ia. This work has made use of data from the European Space Agency (ESA) mission {\it Gaia} (\url{https://www.cosmos.esa.int/gaia}), processed by the {\it Gaia} Data Processing and Analysis Consortium (DPAC, \url{https://www.cosmos.esa.int/web/gaia/dpac/consortium}). Funding for the DPAC has been provided by national institutions, in particular the institutions participating in the {\it Gaia} Multilateral Agreement.

Manuscript written using \texttt{Overleaf}. 
Main analysis performed in \texttt{Python3} \citep{Python3} running on \texttt{Ubuntu} \citep{Ubuntu} systems and \texttt{MS. Windows} running the \texttt{Windows subsystem for Linux (WLS)}.
Radial velocities and selected activity indicators extracted with \texttt{SERVAL} \citep{Zechmeister2018}.
Second RV extraction with \texttt{S-BART} \citep{Silva2022}.
Extensive use of the DACE platform \footnote{https://dace.unige.ch/}
Extensive usage of \texttt{Numpy} \citep{Numpy}.
Extensive usage of \texttt{Scipy} \citep{Scipy}.
All figures built with \texttt{Matplotlib} \citep{Matplotlib}.
Periodograms and phase folded curves built using \texttt{Pyastronomy} \citep{pyastronomy}.
Gaussian processes modelled with \texttt{S+LEAF} \citep{Delisle2022}, \texttt{Celerite2} \citep{Foreman-Mackey2017} and \texttt{George} \citep{Foreman-Mackey2013}.
Nested sampling with \texttt{Dynesty} \citep{Speagle2020}. 
Selected operations accelerated with \texttt{Numba} \citep{Numba}. 
Keplerian function adapted from \texttt{Pytransit} \citep{Parviainen2015}.
Selected calculations made in \texttt{Excel} \citep{Excel}. 
\textit{TESS} target pixel file plot done with \texttt{tpfplotter} \citep{Aller2020}.
Aperture photometry performed with \texttt{AstroImageJ} \citep{Collins2017}.

The bulk of the analysis was performed on desktop PC with an AMD Ryzen$^{\rm TM}$ 9 3950X (16 cores, 2 threads per core, 3.5--4.7 GHz) and a server hosting 2x Intel$^{\rm (R)}$ Xeon$^{\rm (R)}$ Gold 5218 (2x16 cores, 2 threads per core, 2.3--3.9 GHz). 

%
%
\bibliography{biblio}

\begin{appendix}

\section{Other Activity Indicators} \label{append_activity}

\subsection{Chromatic index}

Activity signals created by spots depend on the contrast between the spots and the photosphere of the stars. This contrast is wavelength dependent, larger at blue wavelengths and smaller in the red part of the spectrum. Radial velocity shifts created by spots can create radial velocity trends along the spectral range. The chromatic index is a measure of the slope of that trend in logarithmic wavelength scale across all orders of the echelle spectrographs. We used the chromatic index (CRX) measurements provided by \texttt{SERVAL} for ESPRESSO and CARMENES. Figure~\ref{activity_ind} shows the time series of CRX along with their GLS periodogram. We do not find any significant periodicity in the data.

\subsection{Differential line width}

 We used the differential line width (DLW) measurements provided by \texttt{SERVAL} for ESPRESSO and CARMENES. In the case of some stars it accurately tracks brightness changes of the star in a similar way as high precision photometry can do. This change in DLW due to a change in flux is usually related to the RV trough its gradient \citep{Zicher2022}. Figure~\ref{activity_ind} shows the time series of DLW along with their GLS periodogram. The periodogram shows a peak consistent with the measured stellar rotation.

\subsection{Bisector span of the CCF}

The presence of starspots on the stellar disc distorts the shape of the lines. This effect manifests itself as an asymmetry in the cross correlation function, which is usually related to the gradient of the change in flux caused by said starspots. The scale of the effect is related to the $v \sin i$ of the star, which means that it is not always detectable in slowly rotating stars with small radii. To measure this asymmetry we use the bisector span metric \citep{QuelozHenry2001}.  The bisector span of the CCF in ESPRESSO is automatically provided by the ESPRESSO DRS. Figure~\ref{activity_ind} shows the time series of the bisector span along with its GLS periodogram. We do not find any significant periodicity in the data.

\subsection{Ca II H\&K}

The intensity of the emission of the cores of the Ca II H\&K lines is linked to the strength of the magnetic field of the star, which is well correlated with the rotation period of the star. The measured intensity of the emission also changes when active regions move across the stellar disc, thus tracing the rotation of the star. We calculate the Mount Wilson $S$-index for the ESPRESSO data following \citet{Lovis2011}. We define two triangular-shaped passbands with  full width half maximum (FWHM) of 1.09~{\AA}  centred at 3968.470~{\AA} and 3933.664~{\AA} for the Ca II H\&K line cores, and for the continuum we use two bands 20~{\AA} in width centred at 3901.070~{\AA} (V) and 4001.070~{\AA} (R).
Then the S-index is defined as: 

\begin{equation}
   S=\alpha {{\tilde{N}_{H}+\tilde{N}_{K}}\over{\tilde{N}_{R}+\tilde{N}_{V}}} + \beta,
\end{equation}
\noindent where $\tilde{N}_{H},\tilde{N}_{K},\tilde{N}_{R}$, and $\tilde{N}_{V}$ are the mean fluxes per wavelength unit in each passband,  while $\alpha$ and $\beta$ are calibration constants fixed as $\alpha = 1.111$ and $\beta = 0.0153$ . The S index (S$_{MW}$) serves as a measurement of the Ca II H\&K core flux normalised to the neighbour continuum. Figure~\ref{activity_ind} shows the time series of $S$-index along with their GLS periodogram. This index cannot be computed for the CARMENES data due to its wavelength coverage. We do not find any significant periodicity in the data. 

\subsection{H$\alpha$}

Similar to Ca II H\&K, the emission in the core of the H$\alpha$ line (or filling, in the case of low activity stars) is related to the strength of the magnetic field, and the presence of active regions on the stellar disc. We can also use it to track the motion of said regions aacross the stellar disc, and thus the stellar rotation. We use the H$\alpha$  values provided by SERVAL, which follow the definition of \citet{GomesdaSilva2011}. Figure~\ref{activity_ind} shows the time series of H$\alpha$ along with their GLS periodogram. We do not find any significant periodicity in the data. 

\subsection{LCO Photometry}

\begin{figure*}[ht]
	\includegraphics[width=18cm]{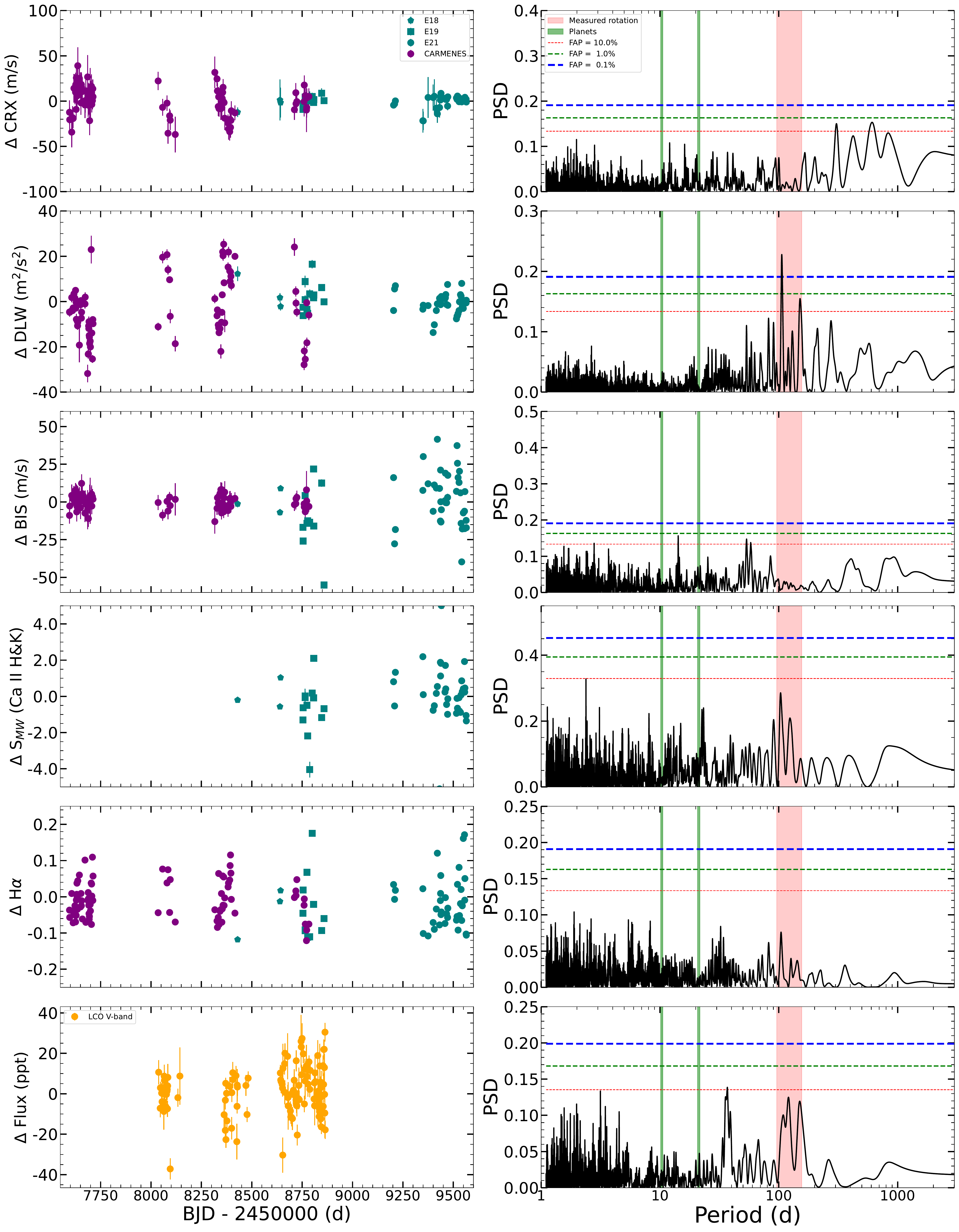}
	\caption{Activity indicators (left panels) and their GLS periodograms (right panels). All spectroscopic indictors have been detrended against the temperature of the \'Echelle grating. The green shaded regions in the GLS periodograms show the periods of the planets GJ~1002 b and c. The widths of the region around the periods have been exaggerated for better visualisation. The red shaded regions shows the 95\% confidence interval around the measured rotation period. In all cases, the median value of each dataset has been subtracted.}
	\label{activity_ind}
\end{figure*}

GJ1002 was observed with the Las Cumbres Observatory (LCO, \citet{Brown2013}) 0.4 m, using the SBIC STL-6303 cameras, from 1st October 2017 to 15 Jan 2020. The raw images were reduced by LCO's pipeline \texttt{BANZAI} and aperture photometry was performed on the calibrated images using \texttt{AstroImageJ} \citep{Collins2017}. For each night a fixed circular aperture was selected by \texttt{AstroImageJ} and aperture photometry was performed using this aperture on the target and a set of 4 reference stars. We obtained 125 nightly binned V-band measurements. The data show an RMS of 11.4 ppt with a typical precision of 4.8 ppt per binned observation. Figure~\ref{activity_ind} shows the time series of V-band photometry along with it GLS periodogram. The periodogram shows no significant periodicities. However, we identify a power excess at around the measured rotation period. 

\section{Parameters of the models}\label{append_tables}

\onecolumn

\fontsize{8}{8}\selectfont
\begin{longtable}{llccccc}
\caption{Priors, measured parameters, and derived parameters of the models described in sections~\ref{analysis_stellar} and ~\ref{analysis_planets} .\label{parameters_1}}\\
\endfirsthead
\multicolumn{7}{c}{\tablename\ \thetable\ -- \textit{Continued from previous page}} \\
\hline
\endhead
\hline \multicolumn{7}{r}{\textit{Continued on next page}} \\
\endfoot
\hline
\endlastfoot
\hline
Parameter  & Priors & 0-Planets & 1-Planet (Circ) &1-Planet (Kep) & 2-Planets (Circ) & 2-Planets (Kep)\\ \hline

\textbf{Zero points} \\
V0 E18$_{FWHM}$ [m~s$^{-1}$] &  
$\mathcal{N}$ (0 , 20) &   
-1.5$^{+4.1}_{-3.8}$ &   
-0.9$^{+4.2}_{-4.2}$ & 
-1.1$^{+4.1}_{-4.3}$ & 
-1.2$^{+3.9}_{-4.0}$ & 
-1.4$^{+3.9}_{-4.0}$\\

V0 E19$_{FWHM}$ [m~s$^{-1}$] &  
$\mathcal{N}$ (0 , 20) &   
-0.3$^{+1.2}_{-1.0}$ & 
-0.3$^{+1.2}_{-1.2}$ &  
-0.2$^{+1.3}_{-1.2}$ & 
-0.8$^{+1.1}_{-1.1}$ & 
-0.8$^{+1.1}_{-1.1}$ \\

V0 E21$_{FWHM}$ [m~s$^{-1}$] &  
$\mathcal{N}$ (0 , 20) &  
0.08$^{+0.70}_{-0.68}$ &  
0.22$^{+0.69}_{-0.70}$ &  
0.18$^{+0.72}_{-0.75}$ & 
0.17$^{+0.66}_{-0.75}$ & 
0.21$^{+0.66}_{-0.71}$ \\
 
V0 CAR$_{FWHM}$ [m~s$^{-1}$] &  
$\mathcal{N}$ (0 , 20) &   
-4.0$^{+2.1}_{-1.9}$ &  
-3.9$^{+2.0}_{-2.0}$ & 
-3.8$^{2.0}_{-2.1}$ & 
-3.7$^{+2.0}_{-2.0}$ &
-3.7$^{+2.0}_{-2.0}$  \\ 

V0 E18$_{RV}$ [m~s$^{-1}$] & 
$\mathcal{N}$ (0 , 5) &   
0.3$^{+1.4}_{-1.4}$ & 
-0.3$^{+1.3}_{-1.2}$ & 
-0.4$^{1.3}_{-1.3}$ & 
0.3$^{+1.5}_{-1.5}$ &  
0.3$^{+1.4}_{-1.5}$\\

V0 E19$_{RV}$ [m~s$^{-1}$] &  
$\mathcal{N}$ (0 , 5) &   
0.66$^{+0.70}_{-0.63}$ &  
0.96$^{+0.72}_{-0.70}$ & 
1.1$^{0.72}_{-0.72}$ & 
0.16$^{+0.65}_{-0.67}$ & 
0.21$^{+0.64}_{-0.64}$\\

V0 E21$_{RV}$ [m~s$^{-1}$] &  
$\mathcal{N}$ (0 , 5) &   
0.36$^{+0.41}_{-0.42}$ &   
0.70$^{+0.45}_{-0.50}$ &
0.68$^{+0.45}_{-0.49}$ & 
0.50$^{+0.55}_{-0.65}$ & 
0.55$^{+0.50}_{-0.61}$\\

V0 CAR$_{RV}$ [m~s$^{-1}$] &  
$\mathcal{N}$ (0 , 5) &   
0.07$^{+0.35}_{-0.39}$ &  
0.11$^{+0.39}_{-0.40}$ & 
0.10$^{0.38}_{0.39}$ & 
0.68$^{+0.47}_{-0.46}$ & 
0.34$^{+0.46}_{-0.45}$\\

\\
\textbf{FWHM vs Temp} \\
Lin ESP [m~s$^{-1}$/mK] &  
$\mathcal{N}$ (0 , 0.2) &   
-0.110$^{+0.064}_{-0.062}$ &  
-0.095$^{+0.066}_{-0.065}$ & 
-0.095$^{+0.067}_{-0.051}$ & 
-0.102$^{+0.064}_{-0.064}$ & 
-0.103$^{+0.064}_{-0.061}$ \\

Quad ESP [m~s$^{-1}$/mK$^{2}$] &  
$\mathcal{N}$ (0 , 0.01) &  
0.0026$^{+0.0051}_{-0.0043}$ & 
0.0024$^{+0.0052}_{-0.0052}$ & 
0.0028$^{+0.0052}_{-0.0051}$ &
0.0022$^{+0.0051}_{-0.0051}$ & 
0.0023$^{+0.0049}_{-0.0050}$\\

Third ESP [m~s$^{-1}$/mK$^{3}$] &  
$\mathcal{N}$ (0 , 0.001) &   
0.00010$^{+0.00015}_{-0.00014}$ &  
0.00010$^{+0.00015}_{-0.00014}$ & 
0.00010$^{+0.00015}_{-0.00014}$ & 
0.00008$^{+0.00015}_{-0.00014}$ & 
0.00008$^{+0.00014}_{-0.00014}$\\

Lin CAR [m~s$^{-1}$/mK] &  
$\mathcal{N}$ (0 , 0.2) &   
-0.116$^{+0.070}_{-0.068}$ &  
-0.122$^{+0.072}_{-0.071}$ &  
-0.122$^{+0.072}_{-0.072}$ &
-0.126$^{+0.072}_{-0.071}$ & 
-0.125$^{+0.072}_{-0.074}$\\

Quad CAR [m~s$^{-1}$/mK$^{2}$] &  
$\mathcal{N}$ (0 , 0.01) &   
0.00218$^{+0.00076}_{-0.00072}$ &  
0.00220$^{+0.00075}_{-0.00069}$ & 
0.00218$^{+0.00073}_{-0.00071}$ &
0.00224$^{+0.00072}_{-0.00070}$ & 
0.00223$^{+0.00069}_{-0.00069}$ \\

Third CAR [m~s$^{-1}$/mK$^{3}$] &  
$\mathcal{N}$ (0 , 0.001) &   
0.000018$^{+0.000015}_{-0.000015}$ &    
0.000018$^{+0.000016}_{-0.000016}$ & 
0.000019$^{+0.000016}_{-0.000016}$ & 
0.000019$^{+0.000016}_{-0.000016}$ & 
0.000019$^{+0.000016}_{-0.000016}$ \\

\\
\textbf{Parameters GP} \\
ln A11$_{FWHM}$ & 
$\mathcal{N}$ (2.3 , 2) &   
-0.72$^{+1.3}_{-1.3}$ &  
-1.07$^{+0.92}_{-1.14}$ & 
-1.05$^{+0.96}_{-1.12}$ & 
-1.48$^{+0.90}_{-1.08}$ & 
-1.46$^{+0.86}_{-1.10}$\\

ln A12$_{FWHM}$ & 
$\mathcal{N}$ (2.3 , 2) &   
3.30$^{+0.47}_{-1.50}$ &  
2.63$^{+0.91}_{-1.88}$ & 
2.67$^{0.88}_{-1.86}$ & 
2.65$^{+0.80}_{-1.77}$ & 
2.69$^{+0.78}_{-1.80}$\\

ln A21$_{FWHM}$ & 
$\mathcal{N}$ (2.3 , 2) &   
0.35$^{+0.62}_{-0.99}$ &  
0.91$^{+0.30}_{-0.49}$ & 
0.91$^{+0.33}_{-0.53}$ & 
0.93$^{+0.27}_{-0.32}$ & 
0.90$^{+0.29}_{-0.36}$  \\

ln A22$_{FWHM}$ & $\mathcal{N}$ (2.3 , 2) &   
0.4$^{+1.1}_{-1.4}$ &  
0.8$^{+1.1}_{-1.4}$ & 
0.7$^{+1.1}_{-1.4}$ & 
0.9$^{+1.1}_{-1.5}$ & 
0.8$^{+1.1}_{-1.4}$\\

ln A11$_{RV}$ & $\mathcal{N}$ (0.8 , 1) & 
-0.19$^{+0.47}_{-0.74}$ & 
0.14$^{+0.30}_{-0.38}$ & 
0.10$^{+0.31}_{-0.39}$ & 
0.45$^{+0.25}_{-0.27}$ & 
0.40$^{+0.26}_{-0.27}$\\

ln A12$_{RV}$ & $\mathcal{N}$ (0.8 , 1) &
0.48$^{+0.84}_{-0.91}$ &  
0.69$^{+0.95}_{-0.98}$ & 
0.64$^{+0.89}_{-0.91}$ & 
0.66$^{+0.92}_{-0.96}$ &  
0.65$^{+0.90}_{-0.94}$\\

ln A21$_{RV}$ & $\mathcal{N}$ (0.8 , 1) &   
0.14$^{+0.47}_{-0.74}$ &  
-0.22$^{+0.48}_{-0.55}$ & 
-0.21$^{+0.51}_{-0.58}$ & 
0.10$^{+0.33}_{-0.44}$ & 
0.08$^{+0.35}_{-0.44}$\\

ln A22$_{RV}$ & $\mathcal{N}$ (0.8 , 1) &   
1.42$^{+0.93}_{-1.17}$ &  
1.26$^{+0.77}_{-0.98}$ & 
1.32$^{+0.71}_{-0.99}$ & 
1.04$^{+0.79}_{-0.96}$ &   
1.09$^{+0.76}_{-0.96}$\\

$P_{\rm rot}$ [d] & $\mathcal{N}$ (130 , 30) &
118$^{+20}_{-15}$ &  
125$^{+23}_{-19}$ & 
125$^{+23}_{-20}$ &   
127$^{+15}_{-14}$ &   
126$^{+15}_{-14}$\\

ln $T_{\rm evol}$ [d] & $\mathcal{N}$ (5.6 , 1) &   
4.47$^{+0.66}_{-0.65}$ &  
4.27$^{+0.61}_{-0.58}$ & 
4.27$^{+0.56}_{-0.57}$ & 
4.28$^{+0.50}_{-0.53}$ &  
4.28$^{+0.50}_{-0.50}$ \\

\\
\textbf{Planet b} \\
$Phase$ & 
$\mathcal{U}$ (0 , 1) &  
& 
& 
& 
0.307$^{+0.021}_{-0.021}$ & 
0.310$^{+0.031}_{-0.029}$\\

$P_{\rm orb}$ [d] & 
$\mathcal{N}$ (10.4 , 0.5) & 
&
& 
& 
10.3465$^{+0.0026}_{-0.0027}$ & 
10.3460$^{+0.0030}_{-0.0027}$\\

$K$ [m~s$^{-1}$] &  
$\mathcal{U}$ (0 , 10) &
& 
& 
& 
1.31$^{+0.14}_{-0.14}$ & 
1.30$^{+0.15}_{-0.15}$\\

$\sqrt{e} ~cos(\omega)$ & 
$\mathcal{N}$ (0 , 0.3) 
& 
& 
& 
& 
& 
-0.03$^{+0.19}_{-0.19}$ \\

$\sqrt{e} ~sin(\omega)$ & 
$\mathcal{N}$ (0 , 0.3) 
& 
& 
& 
& 
& 
-0.01$^{+0.19}_{-0.19}$ \\

M~$\cdot$ sin(i) [M$\oplus$] & 
$Derived$ & 
&
& 
& 
1.08 $\pm$ 0.13 & 
1.06 $\pm$ 0.14\\

$a$ [au] & 
$Derived$ 
&
& 
& 
& 
0.0457 $\pm$ 0.0013 & 
0.0457 $\pm$ 0.0013 \\

$T_{0}$ -- 2450000 [d] &  
$Derived$&  
& 
& 
& 
8583.27 $\pm$ 0.22 & 
8583.30 $\pm$ 0.30 \\

$e$ & 
$Derived$ & 
& 
& 
& 
& 
0.046 $\pm$ 0.07\\

\\
\textbf{Planet c} \\
$Phase$ & 
$\mathcal{U}$ (0 , 1)  &
& 
0.243$^{+0.028}_{-0.027}$ &  
0.206$^{+0.044}_{-0.043}$ & 
0.257$^{+0.024}_{-0.024}$ & 
0.246$^{+0.030}_{-0.033}$ \\

$P_{\rm orb}$ [d] & 
$\mathcal{N}$ (21.2 , 1.0) & 
& 
21.195$^{+0.018}_{-0.017}$ & 
21.186$^{+0.019}_{-0.015}$ & 
21.202$^{+0.013}_{-0.014}$ & 
21.202$^{+0.013}_{-0.013}$ \\ 

$K$ [m~s$^{-1}$] &  
$\mathcal{U}$ (0 , 10) & 
& 
1.20$^{+0.20}_{-0.20}$ & 
1.29$^{+0.23}_{-0.23}$ & 
1.30$^{+0.14}_{-0.14}$ & 
1.30$^{+0.15}_{-0.15}$\\

$\sqrt{e} ~cos(\omega)$ & 
$\mathcal{N}$ (0 , 0.3) & 
& 
& 
0.27$^{+0.21}_{-0.28}$ & 
& 
0.11$^{+0.19}_{-0.20}$ \\

$\sqrt{e} ~sin(\omega)$ & 
$\mathcal{N}$ (0 , 0.3) &
& 
& 
0.11$^{+0.24}_{-0.28}$ & 
& 
0.05$^{+0.20}_{-0.21}$  \\

M ~$\cdot$ sin(i) [M$\oplus$] & 
$Derived$ &
& 
1.27 $\pm$ 0.23 & 
1.31 $\pm$ 0.25 & 
1.36 $\pm$ 0.17 & 
1.36 $\pm$ 0.17 \\

$a$ [au] & 
$Derived$ &
& 
0.0738 $\pm$ 0.0022 &
0.0738 $\pm$ 0.0021 &  
0.0738 $\pm$ 0.0021 & 
0.0738 $\pm$ 0.0021 \\

$T_{0}$ -- 2450000 [d] &  
$Derived$ &  
& 
8585.23 $\pm$ 0.61 &
8584.49 $\pm$ 0.92 & 
8585.54 $\pm$ 0.50 & 
8585.31 $\pm$ 0.64 \\

$e$ & 
$Derived$ & 
& 
& 
0.14 $\pm$ 0.15&
&  
0.06 $\pm$ 0.10\\

\\
\textbf{Jitter} \\
ln Jit E18$_{FWHM}$ & 
$\mathcal{N}$ (2.3 , 2) &   
0.5$^{+1.1}_{-1.4}$ &   
0.5$^{+1.1}_{-1.2}$ & 
0.5$^{+1.1}_{-1.4}$ & 
0.4$^{+1.1}_{-1.4}$ & 
0.4$^{+1.2}_{-1.4}$\\

ln Jit E19$_{FWHM}$ & 
$\mathcal{N}$ (2.3 , 2) &   
0.09$^{+0.70}_{-1.09}$ &   
0.12$^{+0.68}_{-1.18}$ & 
0.12$^{+0.70}_{-1.10}$ & 
0.29$^{+0.63}_{-1.13}$ & 
0.25$^{+0.64}_{-1.11}$ \\

ln Jit E21$_{FWHM}$ & 
$\mathcal{N}$ (2.3 , 2) &
-1.06$^{+0.68}_{-1.01}$ &
-1.15$^{+0.67}_{-1.00}$ & 
-1.09$^{+0.65}_{-1.02}$ & 
-1.14$^{+0.67}_{-1.03}$ & 
-1.13$^{+0.68}_{-1.00}$ \\

ln Jit CAR$_{FWHM}$ & 
$\mathcal{N}$ (2.3 , 2) &   
0.18$^{+0.84}_{-1.25}$ & 
0.06$^{+0.87}_{-1.22}$ & 
0.12$^{+0.84}_{-1.26}$ & 
0.10$^{+0.85}_{-1.23}$ & 
0.14$^{+0.86}_{-1.21}$  \\

ln Jit E18$_{RV}$ & 
$\mathcal{N}$ (0.8 , 1) &   
0.21$^{+0.68}_{-0.69}$ &  
0.03$^{+0.70}_{-0.74}$ & 
-0.02$^{+0.73}_{-0.74}$ &  
0.01$^{+0.74}_{-0.78}$ &  
-0.01$^{+0.70}_{-0.76}$\\

ln Jit E19$_{RV}$ & 
$\mathcal{N}$ (0.8 , 1) &   
0.40$^{+0.39}_{-0.62}$& 
0.47$^{+0.35}_{-0.38}$& 
0.51$^{+0.33}_{-0.39}$& 
-0.43$^{+0.62}_{-0.74}$& 
-0.39$^{+0.62}_{-0.75}$ \\

ln Jit E21$_{RV}$ & 
$\mathcal{N}$ (0.8 , 1) &   
0.03$^{+0.19}_{-0.20}$ & 
-0.07$^{+0.18}_{-0.18}$ & 
-0.11$^{+0.19}_{-0.19}$ &  
-1.08$^{+0.30}_{-0.36}$ &   
-1.04$^{+0.29}_{-0.35}$ \\

ln Jit CAR$_{RV}$ & 
$\mathcal{N}$ (0.8 , 1) &   
0.35$^{+0.20}_{-0.24}$ &  
-0.05$^{+0.27}_{-0.41}$&  
-0.06$^{+0.29}_{-0.42}$ & 
-0.54$^{+0.40}_{-0.54}$ & 
-0.53$^{+0.40}_{-0.57}$\\
\\
\textbf{Model statistics} \\
lnZ & & 
-784.3 & 
-776.9 & 
-777.4 &  
-764.1  & 
-764.5 \\

$\Delta$ lnZ vs 0p & &
& 
7.4 &  
6.9 &  
20.2 & 
19.8  \\

RMS RV [m~s$^{-1}$] & & 1.7 & 1.5 & 1.5 &  1.3 & 1.3 \\
RMS RV$_{ESP}$ [m~s$^{-1}$] & & 0.9 & 0.9 & 0.9 & 0.4 & 0.4 \\
RMS RV$_{CAR}$ [m~s$^{-1}$] & & 2.0 & 1.8 & 1.8 & 1.6 & 1.6 \\
\hline
\end{longtable}

\begin{figure*}
	\includegraphics[width=18cm]{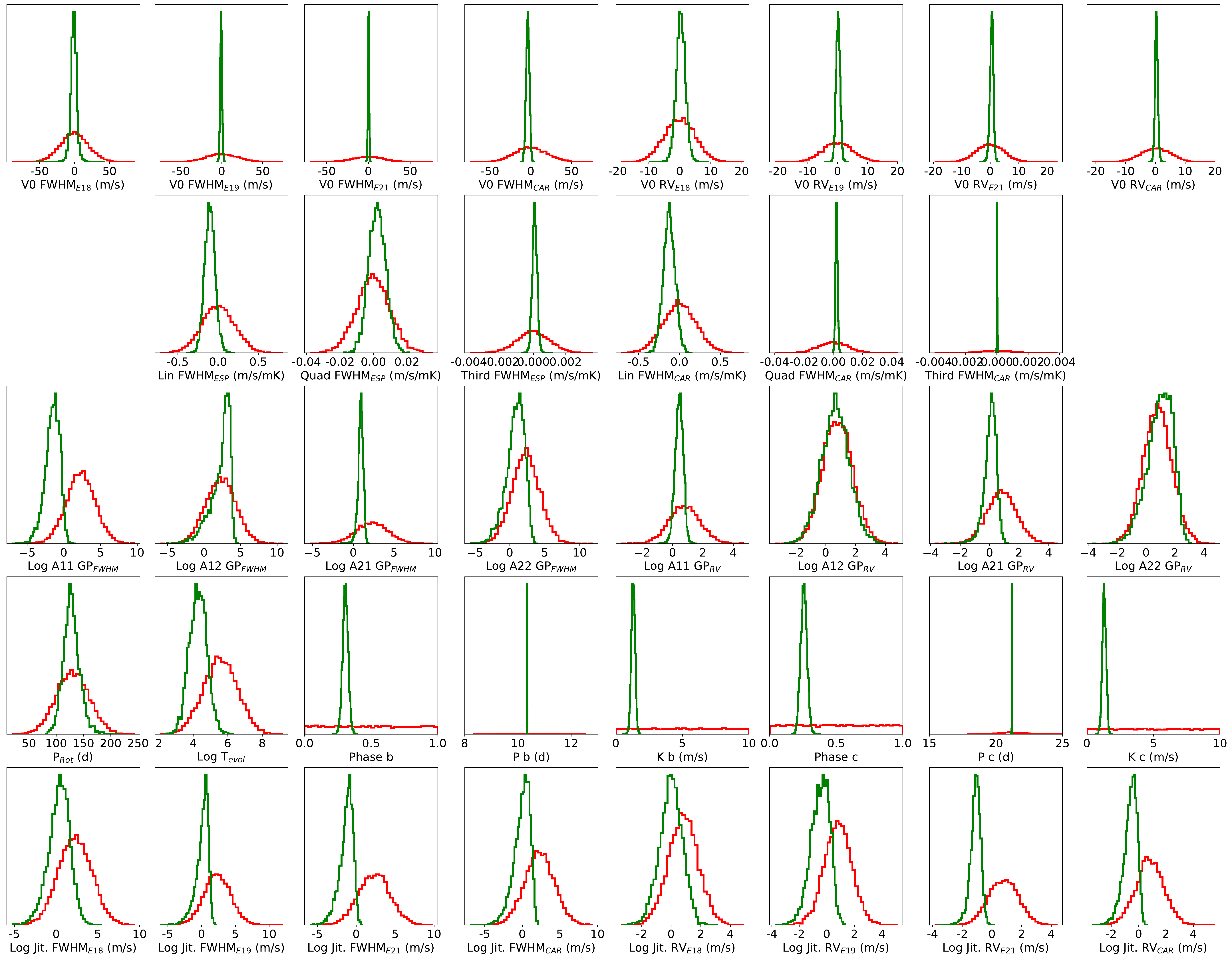}
	\caption{Posterior distribution (green line) of every parameter sampled in the definitive model (2 planets circular). The red line shows the prior distribution in the range of the posterior.}
	\label{posterior_2pc}
\end{figure*}

\label{lastpage}

\end{appendix}

\end{document}